\documentclass[sigconf]{acmart}

\usepackage{graphicx}
\usepackage{textcomp}
\usepackage{subfig}
\usepackage{pifont}
\usepackage{multirow}
\usepackage{multicol}
\usepackage{algorithm}
\usepackage{algpseudocode}
\usepackage{hyperref}
\usepackage{float}
\usepackage{wrapfig}
\usepackage{mwe}
\usepackage{hyperref}
\usepackage{stfloats}
\usepackage{caption}
\usepackage[mathscr]{euscript}
\let \eucal \mathscr
\usepackage{mathrsfs}
\DeclareMathAlphabet{\mathpzc}{OT1}{pzc}{m}{it}
\DeclareMathVersion{normal2}
\usepackage{array}   
\newcolumntype{L}{>{$}l<{$}} 

\copyrightyear{2023}
\acmYear{2023}
\setcopyright{acmlicensed}\acmConference[KDD '23]{Proceedings of the 29th ACM SIGKDD Conference on Knowledge Discovery and Data Mining}{August 6--10, 2023}{Long Beach, CA, USA}
\acmBooktitle{Proceedings of the 29th ACM SIGKDD Conference on Knowledge Discovery and Data Mining (KDD '23), August 6--10, 2023, Long Beach, CA, USA}
\acmPrice{15.00}
\acmDOI{10.1145/3580305.3599803}
\acmISBN{979-8-4007-0103-0/23/08}

\begin{document}

\title{Demystifying Fraudulent Transactions and Illicit Nodes \\ in the Bitcoin Network for Financial Forensics}

\author{Youssef Elmougy}
\affiliation{%
  \institution{Georgia Institute of Technology}
  \city{Atlanta}
  \state{GA 30332}
  \country{USA}
}
\author{Ling Liu}
\affiliation{%
  \institution{Georgia Institute of Technology}
  \city{Atlanta}
  \state{GA 30332}
  \country{USA}
}

\renewcommand{\shortauthors}{Youssef Elmougy and Ling Liu}

\begin{abstract}
Blockchain provides the unique and accountable channel for financial forensics by mining its open and immutable transaction data. A recent surge has been witnessed by training machine learning models with cryptocurrency transaction data for anomaly detection, such as money laundering and other fraudulent activities. This paper presents a holistic applied data science approach to fraud detection in the Bitcoin network with two original contributions. 
First, we contribute the \textit{Elliptic++ dataset}, which extends the Elliptic transaction dataset to include over $822k$ Bitcoin wallet addresses (nodes), each with $56$ features, and $1.27M$ temporal interactions. This enables both the detection of fraudulent transactions and the detection of illicit addresses (actors) in the Bitcoin network by leveraging four types of graph data: (i) the transaction-to-transaction graph, representing the money flow in the Bitcoin network, (ii) the address-to-address interaction graph, capturing the types of transaction flows between Bitcoin addresses, (iii) the address-transaction graph, representing the bi-directional money flow between addresses and transactions (BTC flow from input address to one or more transactions and BTC flow from a transaction to one or more output addresses), and (iv) the user entity graph, capturing clusters of Bitcoin addresses representing unique Bitcoin users. 
Second, we perform fraud detection tasks on all four graphs by using diverse machine learning algorithms. We show that adding enhanced features from the address-to-address and the address-transaction graphs not only assists in effectively detecting both illicit transactions and illicit addresses, but also assists in gaining in-depth understanding of the root cause of money laundering vulnerabilities in cryptocurrency transactions and the strategies for fraud detection and prevention. The Elliptic++ dataset is released at \url{https://www.github.com/git-disl/EllipticPlusPlus}. 
\end{abstract}



\maketitle

\vspace{-0.2cm}
\section{Introduction}
The emergence of cryptocurrency, initiated by Bitcoin~\cite{nakamoto2008bitcoin}, has kindled an adoption of a new exchange system in which financial transactions can be completed without an intermediary. This popularity has been perceived as a catalyst for the integration of cryptocurrencies into the payment systems of traditional financial institutions, enabling processing of exchanges between fiat money and cryptocurrency. 
To allow for immense volumes of such exchanges, financial institutions must provide rigid security against high risk transactions from fraudulent activities and assure compliance with anti-money laundering (AML) regulations. 
Although machine learning (ML) models have become a popular tool for anomaly detection and risk assessment of cryptocurrency transactions in recent years, we argue that fraud detection models trained for financial forensics 
should revolve around two goals: (i) prioritizing higher recall with a (reasonable) trade-off on precision, since the penalty of classifying an illicit transaction/account as licit far outweighs that of the reverse, and (ii) allowing explainability (and derivation of proof) on the risk of transactions/accounts.

Bitcoin~\cite{nakamoto2008bitcoin} is the most widely used cryptocurrency.
Analyzing its blockchain will provide a basis of reference. The Elliptic dataset~\cite{weber2019anti} is 
the largest labelled Bitcoin transaction data publicly available.
The dataset consists of over $203k$ transactions labelled illicit, licit, and unknown. 
It provides a good entry into fraudulent transaction analysis. However, the Elliptic dataset~\cite{weber2019anti} consists of only Bitcoin transactions, without features of the addresses involved and the different interactions between pairs of addresses. Hence, the forensic analysis using the Elliptic dataset suffers from 
a prominent downfall: when a model predicts an illicit transaction, the addresses responsible cannot be clearly identified since a transaction may be associated with several input and output addresses.

The first contribution of this paper is the \textit{Elliptic++ dataset}, which extends the Elliptic dataset to include all the Bitcoin wallet addresses (actors) and their temporal interactions associated to the transactions in the Elliptic dataset. Hence, forensics performed using the Elliptic++ dataset can identify fraudulent transactions and dishonest actors in the Bitcoin network and will also allow explainability as to why a wallet address associated with a transaction is illusive and dishonest. 
The construction of the \textit{Elliptic++ dataset} contributes a novel way in collecting and visualizing Blockchain data, using wallet addresses as the center of a risk detection model. We collect Blockchain data associated with the Elliptic dataset (transactions). We first augment each transaction in the \textit{transactions dataset} with $17$ additional features, then we crawl the Blockchain to create the Elliptic++ dataset, consisting of the \textit{feature-enhanced transactions dataset} and the \textit{actors (wallet addresses) dataset}. The actors dataset includes over $822k$ labelled wallet addresses, each described with 56 features, and over $1.27M$ temporal occurrences (interactions) across the same time steps (as those recorded in the Elliptic dataset). With our Elliptic++ dataset, one can perform anomaly detection of fraudulent activities, such as illicit transactions and fraudulent actors. We also include four types of graphs: the \textit{Money Flow Transaction Graph}, the \textit{Actor Interaction Graph}, the \textit{Address-Transaction Graph}, and the \textit{User Entity Graph}. These graphs contribute to both mining and visualization of the connections of transactions and the interactions among wallet addresses through their transactions.

The second contribution of the paper is   
performing fraud detection using the Elliptic++ dataset and the four graph representations by combining diverse ML algorithms and feature optimizations. 
We observe that Random Forest (RF) with feature refinement offers the best-performing model: on the transactions dataset, it achieves $98.6\%$ precision and $72.7\%$ recall compared to $97.5\%$ precision and $71.9\%$ recall when using RF without feature refinement; on the actors dataset, RF with feature refinement achieves $92.1\%$ precision and $80.2\%$ recall, compared to $91.1\%$ precision and $78.9\%$ recall when using RF without feature refinement. Furthermore, the fraud detection using the Elliptic++ dataset allows for in-depth understanding of the root cause of fraudulent activities in cryptocurrency transactions through semantic and statistical explainability, shining light on the strategies for fraud detection and prevention.




\vspace{-0.3cm}
\section{Related Work}
Since the introduction of the Elliptic dataset in 2019 by Weber et al.~\cite{weber2019anti}, there have been numerous efforts on data labelling and anomaly detection using ML models. Lorenz et al.~\cite{lorenz2020machine} assumed minimal access to labels, and proposes an active learning solution by leveraging a smaller subset of the available labels. Oliveira et al.~\cite{oliveira2021guiltywalker} proposed GuiltyWalker, a detection method that computes new features based on the structure of a transaction-to-transction graph and the distance to known illicit transactions. Alarab et al.~\cite{alarab2020comparative} 
presented a comparative analysis of the Elliptic dataset using different supervised learning methods. Loa et al.~\cite{loainspection} proposed Inspection-L, a graph neural network framework for anomaly detection, and similarly, Alarab et al.~\cite{alarab2022graph} proposed a graph-based LSTM with a graph convolutional network to detect illicit transactions.



More generally, there is increasing interest in AML in the context of financial transactions and cryptocurrency networks, such as Bitcoin or Ethereum. Combinatorial optimization methods for identifying transactions from graphs, including statistical deviation and dense subgraph discovery methods, have been explored in~\cite{sun2022monlad, chen2023heavy}. The lack of labelled data and the imbalanced data are two major challenges for fraud detection when using supervised learning~\cite{bartoletti2018data, hu2019characterizing, monamo2016multifaceted} or unsupervised learning~\cite{monamo2016unsupervised, monamo2016multifaceted, pham2016anomaly, pham2016anomaly2, hirshman2013unsupervised}. 
Some efforts have also focused on de-anonymizing mixed transactions by Bitcoin-dedicated graph-based approaches~\cite{biryukov2014deanonymisation, koshy2014analysis, biryukov2019deanonymization, wu2021detecting}. 
Regarding Bitcoin account profiling, 
Michalski et al.~\cite{michalski+2020-IEEEaccess} built a small dataset of $9,000$ addresses and applied supervised learning to characterize nodes in the Blockchain as miner or exchange, indicating the need of countermeasures for preserving the desired level of anonymity. To the best of our knowledge, our Elliptic++ dataset is the largest public Bitcoin account dataset with $822,942$ addresses. 

Existing works on Ethereum account profiling and phishing account detection exclusively focus on graph representation learning methods~\cite{Sihao+WWW2023,Sihao+PSBert-2023}. 
For Ethereum account profiling, \cite{shen2021identity} and \cite{zhou2022behavior} applied either Graph Convolution Network (GCN)~\cite{gcn} or a hierarchical graph attention encoder (HGATE) to infer the identity of Ethereum accounts based on learning latent features using node-embedding and subgraph-embeddings. TTAGNN~\cite{li2022ttagn} generate node embedding by combining LSTM encoder and Graph Attention Network (GAT)~\cite{gat}. 
For detecting phishing accounts, Trans2Vec~\cite{trans2vec} and its variants~\cite{lin2020modeling,lin2020t} utilize a graph based random walk algorithm on transaction time and amount information. The cascade method~\cite{chen2020phishing} leverages statistical feature extraction and a lightGBM-based ensemble algorithm. It is worth to note that given the difference in fundamental settings (UTXO and account models), it is not straightforward to apply the existing graph representation learning models developed for profiling Ethereum accounts to profiling Bitcoin accounts on the Bitcoin network. 

Compared to the literature, this paper presents novel contributions from two perspectives. First, it describes and makes publicly available a labelled dataset with Bitcoin blockchain transactions and wallet addresses. The main benefit of this dataset is that it allows more accurate fraud detection models and algorithms to be developed. Second, it uses the 
dataset to showcase the detection of illicit blockchain transactions, illicit actors, and the risks of de-anonymizing users based on address clustering. For 
this, it utilizes a number of baseline machine learning models.

\begin{table*}
\renewcommand{\arraystretch}{1.3}
\centering
\caption{Data structure for an example transaction node in the transactions dataset: (1) \textit{features}, row of all $183$ feature values for \texttt{txId}, (2) \textit{edgelist}, all incoming and outgoing edges involving \texttt{txId}, and (3) \textit{classes}, the class label for \texttt{txId}.}
\vspace{-0.5cm}
\label{tab:onetxsexample}
\begin{tabular}{ccccccccccc}
\multicolumn{11}{c}{\texttt{txs\_features.csv}} \\ \hline
\multicolumn{1}{|c|}{txId} & \multicolumn{1}{c|}{Time step} & \multicolumn{1}{c|}{LF\_1} & \multicolumn{1}{c|}{$\cdots$} & \multicolumn{1}{c|}{LF\_93} & \multicolumn{1}{c|}{AF\_1} & \multicolumn{1}{c|}{$\cdots$} & \multicolumn{1}{c|}{AF\_72} &  \multicolumn{1}{c|}{TXS\_in} & \multicolumn{1}{c|}{$\cdots$} & \multicolumn{1}{c|}{BTC\_out\_total}\\ \hline
\multicolumn{1}{|c|}{\textbf{272145560}} & \multicolumn{1}{c|}{24} & \multicolumn{1}{c|}{-0.155493} & \multicolumn{1}{c|}{$\cdots$} & \multicolumn{1}{c|}{1.135279} & \multicolumn{1}{c|}{-0.159681} & \multicolumn{1}{c|}{$\cdots$} & \multicolumn{1}{c|}{1.521399} & \multicolumn{1}{c|}{1} & \multicolumn{1}{c|}{$\cdots$} & \multicolumn{1}{c|}{2.77279994} \\ \hline
 &  &  &  &  &  &  &  \\[-3ex]
\textbf{} & \multicolumn{2}{c}{\texttt{txs\_edgelist.csv}} & \textbf{} & \multicolumn{2}{c}{\texttt{txs\_classes.csv}} &  &  \\ \cline{2-3} \cline{5-6}
\multicolumn{1}{c|}{} & \multicolumn{1}{c|}{txId1} & \multicolumn{1}{c|}{txId2} & \multicolumn{1}{c|}{} & \multicolumn{1}{c|}{txId} & \multicolumn{1}{c|}{class} &  &  \\ \cline{2-3} \cline{5-6}
\multicolumn{1}{c|}{\textbf{}} & \multicolumn{1}{c|}{\textbf{272145560}} & \multicolumn{1}{c|}{296926618} & \multicolumn{1}{c|}{} & \multicolumn{1}{c|}{\textbf{272145560}} & \multicolumn{1}{c|}{1} &  &  \\ \cline{2-3} \cline{5-6}
\multicolumn{1}{c|}{\textbf{}} & \multicolumn{1}{c|}{\textbf{272145560}} & \multicolumn{1}{c|}{272145556} &  &  &  &  &  \\ \cline{2-3}
\multicolumn{1}{c|}{} & \multicolumn{1}{c|}{299475624} & \multicolumn{1}{c|}{\textbf{272145560}} &  &  &  &  &  \\ \cline{2-3}
\end{tabular}
\vspace{-0.1cm}
\end{table*}

\section{The Elliptic++ Dataset} \label{datasetintro}
The Elliptic++ dataset consists of $203k$ Bitcoin transactions and $822k$ wallet addresses. It leverages elements from the Elliptic dataset\footnote{\url{www.kaggle.com/datasets/ellipticco/elliptic-data-set}}~\cite{weber2019anti}, a published dataset deanonymizing $99.5\%$ of Elliptic transaction data\footnote{\url{www.kaggle.com/datasets/alexbenzik/deanonymized-995-pct-of-elliptic-transactions}}, and the Bitcoin addresses dataset obtained by using our Bitcoin Blockchain\footnote{\url{www.blockchain.com}} scraping pipeline. A detailed description of the data collection pipeline is included in the Appendix.

\vspace{-0.2cm}
\subsection{Transactions Dataset}
\begin{figure}
\centering{
\vspace{-0.4cm}
\caption{Number of transactions by time step.}
\label{fig:txstimesteps}
\makebox[\linewidth]{\includegraphics[width=3.5in]{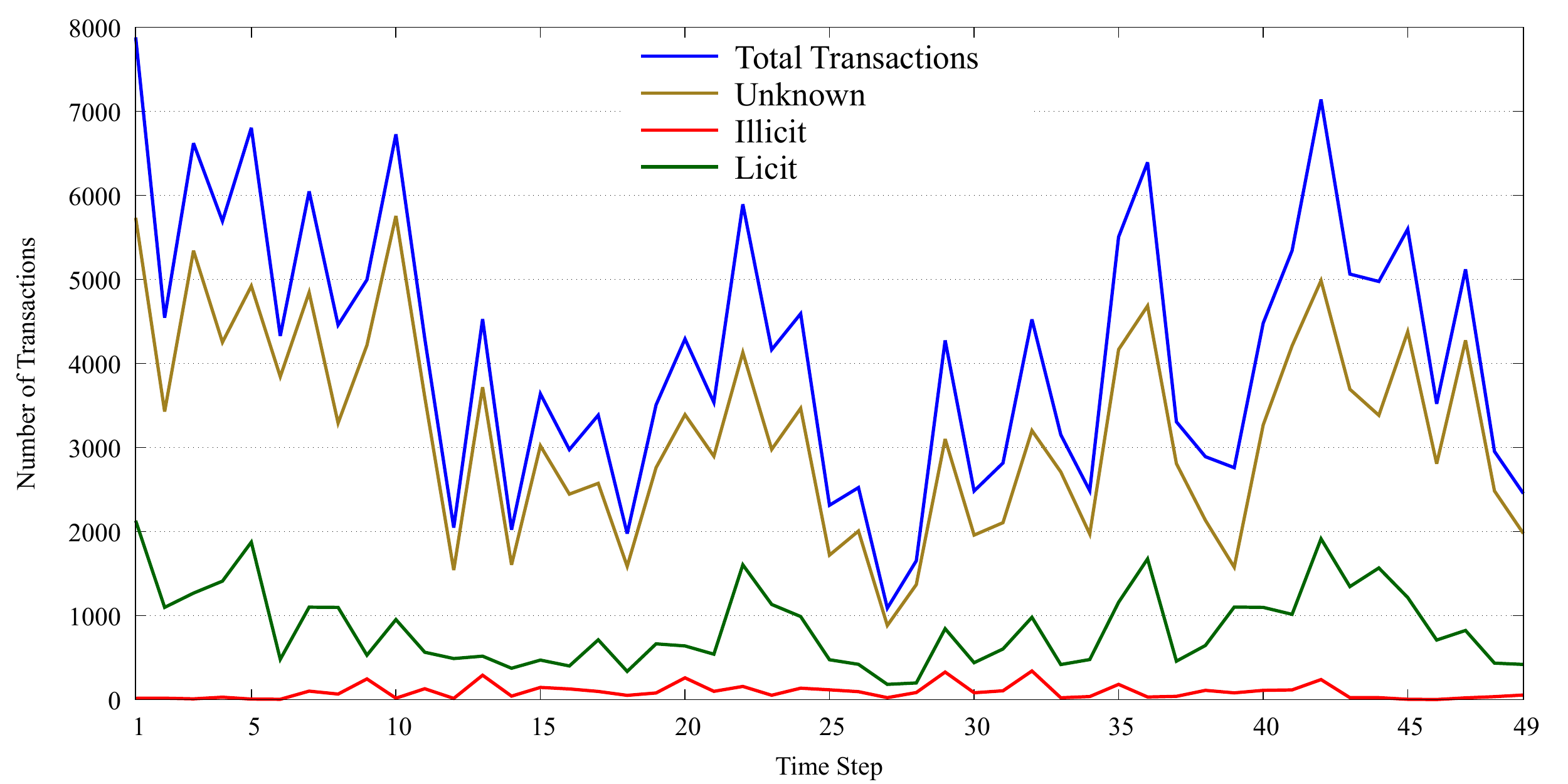}\hspace{0.2cm}}
}
\vspace{-0.7cm}
\end{figure}

The \textit{transactions dataset} consists of a time-series graph with $49$ distinct time steps, $203,769$ transactions (nodes), and $234,355$ directed edges representing the payment flows. Each transaction node is labelled as illicit, licit, or unknown; with $2\%$ ($4,545$) labelled as class-$1$ (illicit), $21\%$ ($42,019$) as class-$2$ (licit), and the remaining transactions are unknown with regard to licit/illicit, hence labelled as class-$3$. Three \texttt{csv} files are used, as shown in Table~\ref{tab:onetxsexample}. Each transaction node has an entry in \texttt{txs\_features.csv}, with numerical data for $183$ transaction features, and an entry in \texttt{txs\_classes.csv}, representing its class label ($1$: \textit{illicit}, $2$: \textit{licit}, $3$: \textit{unknown}). Each edge has an entry in \texttt{txs\_edgelist.csv} (indexed by two transaction IDs), representing money flow from one transaction to another. 
Among the $183$ node features, $166$ features are inherited from the Elliptic dataset, i.e., the time step, $93$ \textit{local features}, representing local information about the transaction, and $72$ \textit{aggregate features}, obtained by aggregating transaction information one-hop forward/backward. The remaining $17$ node features are gathered by our Elliptic++ data collection pipeline (with the exception of the $0.5\%$ of transactions that were not deanonymized) as \textit{augmented features} and are shown in Table~\ref{tab:augmentedfeatures}.  
Figure~\ref{fig:txstimesteps} shows the distribution of transactions across the three classes in each of the 49 time steps.

\begin{table}[t!]
\renewcommand{\arraystretch}{1.3}
\centering
\caption{Transactions dataset augmented features.}
\label{tab:augmentedfeatures}
\vspace{-0.3cm}
\begin{tabular}{p{0.35\linewidth} p{0.6\linewidth}}
\hline
Feature & Description \\ \hline
$BTC_{in}$ & Total BTC incoming\\
$BTC_{out}$ & Total BTC outgoing\\\cline{2-2} 
\multicolumn{1}{r}{\textbf{each has 5 values: }} & \textbf{total, min, max, mean, median} \\ \hline
$Txs_{in}$ & Number of incoming transactions\\
$Txs_{out}$ & Number of outgoing transactions\\
$Addr_{in}$ & Number of input addresses\\
$Addr_{out}$ & Number of output addresses\\
$BTC_{total}$ & Total BTC transacted\\
$Fees$ & Total fees in BTC\\
$Size$ & Total transaction size\\\cline{2-2} 
 & \textbf{single value} \\ \hline
\end{tabular}
\vspace{-0.35cm}
\end{table}

\begin{table}[t!]
\renewcommand{\arraystretch}{1.3}
\centering
\caption{Actors (wallet addresses) dataset features.}
\label{tab:addressfeatures}
\vspace{-0.3cm}
\begin{tabular}{p{0.35\linewidth} p{0.6\linewidth}}
\hline
Feature & Description \\ \hline
\textbf{Transaction related}: &  \\ \cline{1-1}
$BTC_{transacted}$ & Total BTC transacted (sent+received)\\
$BTC_{sent}$ & Total BTC sent\\
$BTC_{received}$ & Total BTC received\\
$Fees$ & Total fees in BTC\\
$Fees_{share}$ & Total fees as share of BTC transacted\\
\textbf{Time related}: &  \\ \cline{1-1}
$Blocks_{txs}$ & Number of blocks between transactions\\
$Blocks_{input}$ & Number of blocks between being an input address\\
$Blocks_{output}$ & Number of blocks between being an output address\\
$Addr\;interactions$ & Number of interactions among addresses\\ \cline{2-2} 
\multicolumn{1}{r}{\textbf{each has 5 values: }} & \textbf{total, min, max, mean, median} \\ \hline
$Class$ & class label: \{illicit, licit, unknown\}\\
\textbf{Transaction related}: &  \\ \cline{1-1}
$Txs_{total}$ & Total number of blockchain transactions\\
$Txs_{input}$ & Total number of dataset transactions as input address\\
$Txs_{output}$ & Total number of dataset transactions as output address\\
\textbf{Time related}: &  \\ \cline{1-1}
$Timesteps$ & Number of time steps transacting in\\
$Lifetime$ & Lifetime in blocks\\
$Block_{first}$ & Block height first transacted in\\
$Block_{last}$ & Block height last transacted in\\
$Block_{first\;sent}$ & Block height first sent in\\
$Block_{first\;receive}$ &  Block height first received in\\
$Repeat\;interactions$ & Number of addresses transacted with multiple times\\ \cline{2-2} 
 & \textbf{single value} \\ \hline
\end{tabular}
\vspace{-0.3cm}
\end{table}

\begin{table*}
\renewcommand{\arraystretch}{1.3}
\centering
\caption{Data structure for an example address in the actors dataset: (1) \textit{features}, row of $56$ feature values for \texttt{address}, (2) \textit{classes}, the class label for \texttt{address}, (3) \textit{AddrAddr}, all edges involving \texttt{address} in the actor interaction graph, (4) \textit{AddrTx}, all edges as involving \texttt{address} as the input address in the address-transaction graph, and similarly (5) \textit{TxAddr}, as the output address.}
\vspace{-0.5cm}
\label{tab:oneaddrexample}
\makebox[\linewidth]{
\begin{tabular}{ccccccc}
\multicolumn{7}{c}{\texttt{wallets\_features.csv}} \\ \hline
\multicolumn{1}{|c|}{address} & \multicolumn{1}{c|}{time step} & \multicolumn{1}{c|}{txs\_input} & \multicolumn{1}{c|}{$\cdots$}  &  \multicolumn{1}{c|}{lifetime\_blocks} & \multicolumn{1}{c|}{$\cdots$} & \multicolumn{1}{c|}{Addr\_interactions\_median} \\ \hline
\multicolumn{1}{|c|}{\textbf{39sfuA8pY4UfybgEZi7uvA13jkGzZpsg5K}} & \multicolumn{1}{c|}{23} & \multicolumn{1}{c|}{420} & \multicolumn{1}{c|}{$\cdots$} & \multicolumn{1}{c|}{18145} & \multicolumn{1}{c|}{$\cdots$} &\multicolumn{1}{c|}{1} \\ \hline
\multicolumn{2}{c}{\texttt{AddrAddr\_edgelist.csv}} & & \multicolumn{2}{c}{\texttt{wallets\_classes.csv}}  & &\\ \cline{1-2} \cline{4-5}
\multicolumn{1}{|c|}{input\_address} & \multicolumn{1}{c|}{output\_address} &  \multicolumn{1}{c|}{} & \multicolumn{1}{c|}{address} & \multicolumn{1}{c|}{class}  & &\\ \cline{1-2} \cline{4-5}
\multicolumn{1}{|c|}{\textbf{39sfuA8pY4UfybgEZi7uvA13jkGzZpsg5K}} & \multicolumn{1}{c|}{1ML...kTL} & \multicolumn{1}{c|}{} & \multicolumn{1}{c|}{\textbf{39sf...sg5K}} &\multicolumn{1}{c|}{1} & &\\ \cline{1-2} \cline{4-5}
\multicolumn{2}{c}{\texttt{AddrTx\_edgelist.csv}} &  & \multicolumn{4}{c}{\texttt{TxAddr\_edgelist.csv}}\\ \cline{1-2} \cline{4-7}
\multicolumn{1}{|c|}{input\_address} & \multicolumn{1}{c|}{txId} &    & \multicolumn{1}{|c|}{txId} & \multicolumn{3}{c|}{output\_address} \\ \cline{1-2} \cline{4-7} 
\multicolumn{1}{|c|}{\textbf{39sfuA8pY4UfybgEZi7uvA13jkGzZpsg5K}} & \multicolumn{1}{c|}{272145560} & & \multicolumn{1}{|c|}{322554634} & \multicolumn{3}{c|}{\textbf{39sfuA8pY4UfybgEZi7uvA13jkGzZpsg5K}} \\ \cline{1-2} \cline{4-7}
\end{tabular}}
\vspace{-0.2cm}
\end{table*}

\begin{figure}
\vspace{-0.4cm}
\centering{
\caption{Number of wallet addresses by time step.}
\label{fig:walletstimesteps}
\makebox[\linewidth]{\includegraphics[width=3.5in]{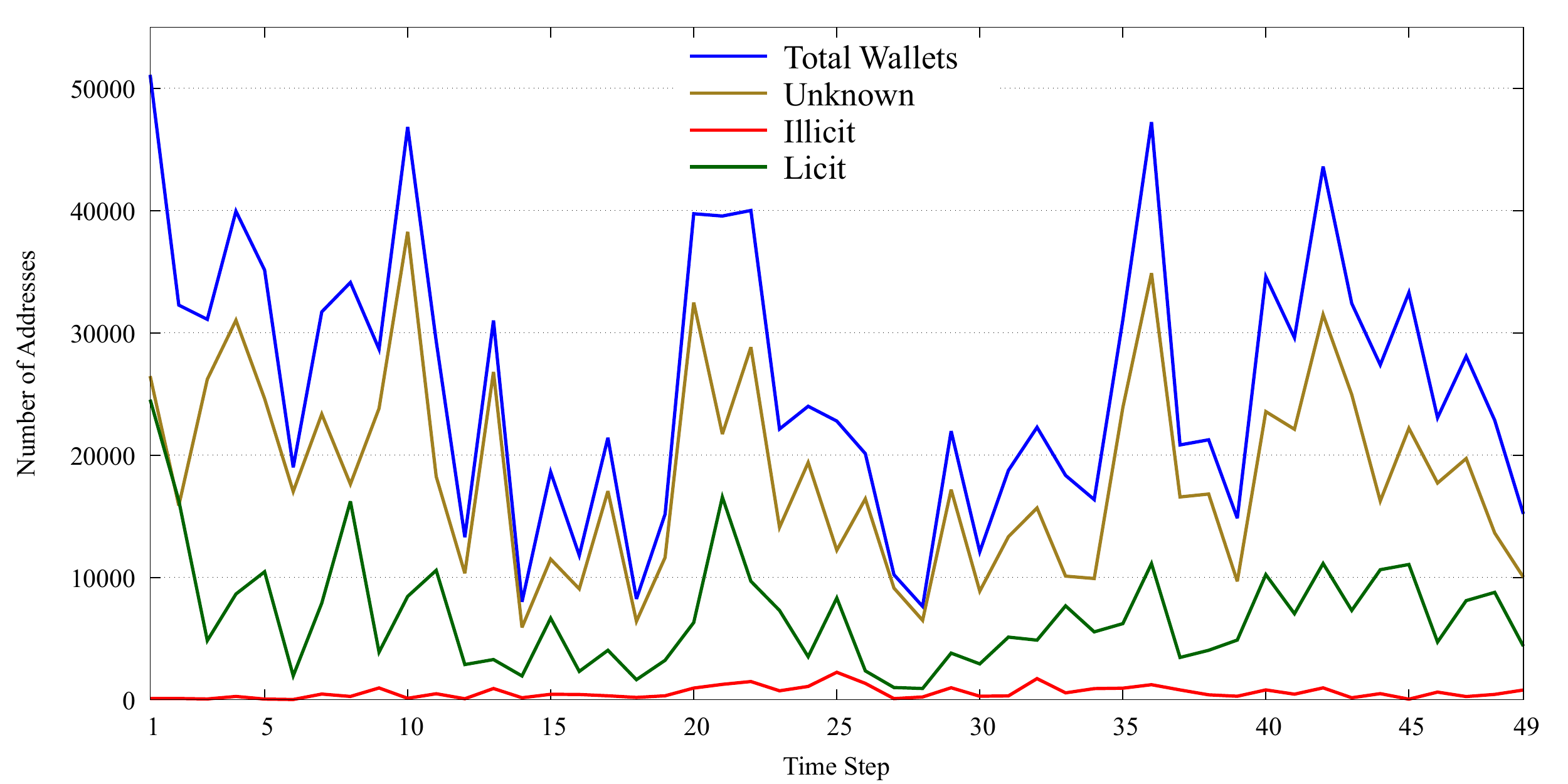}\hspace{0.2cm}}
}
\vspace{-1cm}
\end{figure} 

\subsection{Actors (Wallet Addresses) Dataset}
The \textit{actors (wallet addresses) dataset} is a graph network of $822,942$ wallet addresses, each with $56$ features as shown in Table~\ref{tab:addressfeatures}. 
Five \texttt{csv} files are used, as shown in Table~\ref{tab:oneaddrexample}. 
Each address has an entry in \texttt{wallets\_features.csv}, with numerical data for the time step and $56$ address features listed in Table~\ref{tab:addressfeatures} for each transaction it was involved in, and an entry in \texttt{wallets\_classes.csv}, representing its class label ($1$: \textit{illicit}, $2$: \textit{licit}, $3$: \textit{unknown}). 
A wallet address is labelled {\it illicit} if it has at least $1$ edge with an illicit transaction, otherwise it is labelled {\it licit} if the ratio of its total unknown transactions (edges) to its total licit transactions (edges) is $>3.7$ or {\it unknown} if $\leq 3.7$. The ratio "$3.7$" is calculated using the total ratio of unknown transactions to licit transactions in the transactions dataset. 
We also create \texttt{AddrAddr\_edgelist.csv} to record the pairwise interactions of input and output addresses through Bitcoin transactions. Each entry represents the input address and output address relationship of one transaction. If there are multiple transactions between a pair of addresses, then there are multiple entries in this table. 
Additionally, we create \texttt{AddrTx\_edgelist.csv}, where each entry represents the relationship between an input address and a transaction, and \texttt{TxAddr\_edgelist.csv}, with each entry representing a directed connection between a transaction and an output address. With these data structures, the Elliptic++ dataset can be used to build the address-to-address graph (Section~\ref{actorinteractionsection}), 
and the address-transaction graph (Section~\ref{addrtxaddr}), in addition to the transaction-to-transaction money flow graph (Section~\ref{moneyflowsection}).

The distribution of address classes are $2\%$ ($14,266$) class-$1$ (illicit), $31\%$ ($251,088$) class-$2$ (licit), and the remaining are unknown (class-$3$). When populated using temporal information provided by the time steps in the transactions dataset, we obtain $1,268,260$ wallet address occurrences across all time steps, including $2\%$ ($28,601$) illicit addresses, $27\%$ ($338,871$) licit addresses, and $71\%$ ($900,788$) unknown addresses, as shown in Figure~\ref{fig:walletstimesteps}. 

\subsection{Graph Visualization} \label{graphviz}
In this section, we provide a detailed explanation with visualization of the Elliptic++ dataset using four types of graphs: Money Flow Transaction Graph, Actor Interaction Graph, Address-Transaction Graph, and User Entity Graph.  Each graph has its own unique importance.
\begin{itemize}
\item The \textbf{\textit{Money Flow Transaction Graph}} shows BTC flow from one transaction to the next, allowing exploration of the spatial and temporal patterns surrounding a given transaction.
\item The \textbf{\textit{Actor Interaction Graph}} shows the pairwise interactions among input and output addresses of transactions, showing the density of the k-hop neighborhoods of wallet addresses.
\item The \textbf{\textit{Address-Transaction Graph}} is a heterogenous graph showing flow of BTC across transactions and addresses, allowing an evaluation of purposes of a transaction and the relationships among addresses of the same transaction.
\item The \textbf{\textit{User Entity Graph}} is an address cluster graph that allows for potential linking of addresses controlled by a specific user, which further de-anonymizes their identity and purpose.
\end{itemize}





\subsubsection{Money Flow Transaction Graph}\label{moneyflowsection}
\begin{figure}
\vspace{-0.6cm}
\centering{
\caption{Money flow transaction graphs for selected unknown (left), illicit (middle), and licit (right) transactions in time step $32$ (top shows distribution of TS $32$ transactions).}
\label{fig:moneyflowalltxs}
 \hspace{-0.4cm}
\makebox[\linewidth]{\includegraphics[width=3.5in]{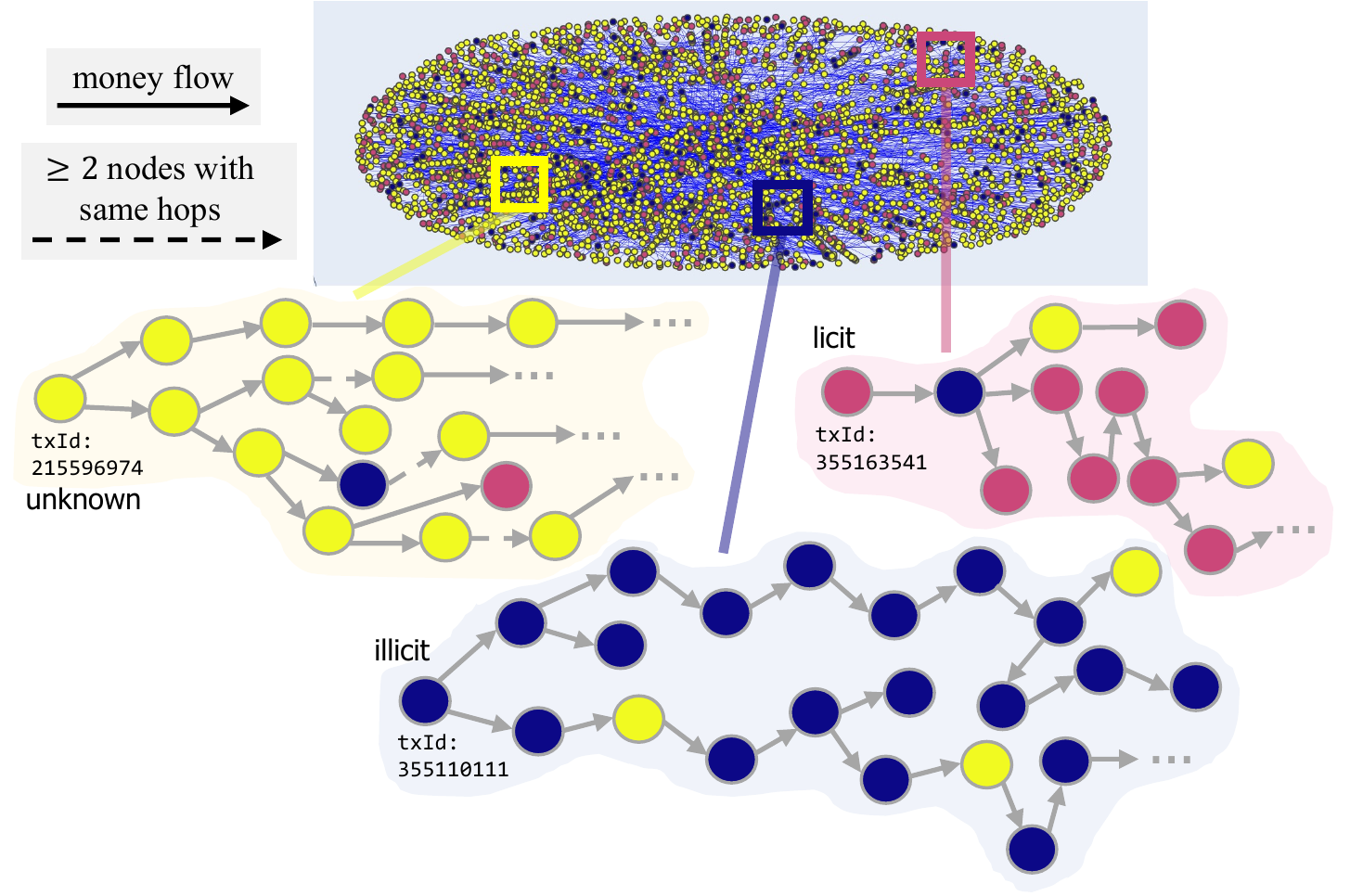}}
}
\vspace{-0.5cm}
\end{figure}

This is a directed graph where nodes represent transactions and edges represent directed BTC flows from one transaction to the next. 
Figure~\ref{fig:moneyflowalltxs} visualizes the distribution of all transactions in time step $32$. 
We choose three transactions (one per class) representing the unknown (yellow), illicit (blue), and licit (red) classes, and display sub-graphs of their four-hop neighbourhoods as shown in the left, middle, and right of Figure~\ref{fig:moneyflowalltxs}. 
All neighbour nodes are shown, though for visual clarity, some edges are abbreviated with dotted arrows signifying that a particular node has two (or more) outgoing edges of the same pattern as the graph is traversed further. 
This displays the potential utility of exploring the spatial and temporal patterns surrounding a given transaction. 

It is important to note that some time steps produce sparse transaction graphs, e.g. time step $14$, while others produce dense transaction graphs, e.g. time step $32$. We provide a comparison in the Appendix (see Figure~\ref{fig:txstimestep1432}). This difference in graph density drastically affects the node neighborhood patterns, with some time steps creating shallow, wide money flow transaction graphs, while other time steps creating deep, narrow money flow transaction graphs. 
Moreover, although there are only $2\%$ illicit transactions within the dataset, they are evidently spread out across each time step and hence the ML model trained for anomaly detection over the earlier sequence of time steps can be leveraged to perform forensics on the later part of the sequence of time steps. 

\subsubsection{Actor Interaction Graph} \label{actorinteractionsection}
\begin{figure}[h]
\vspace{-0.4cm}
\centering{
\caption{Distribution of wallet addresses in time step $32$.}
\label{fig:walletstimestep32}
\makebox[\linewidth]{\includegraphics[width=3.5in]{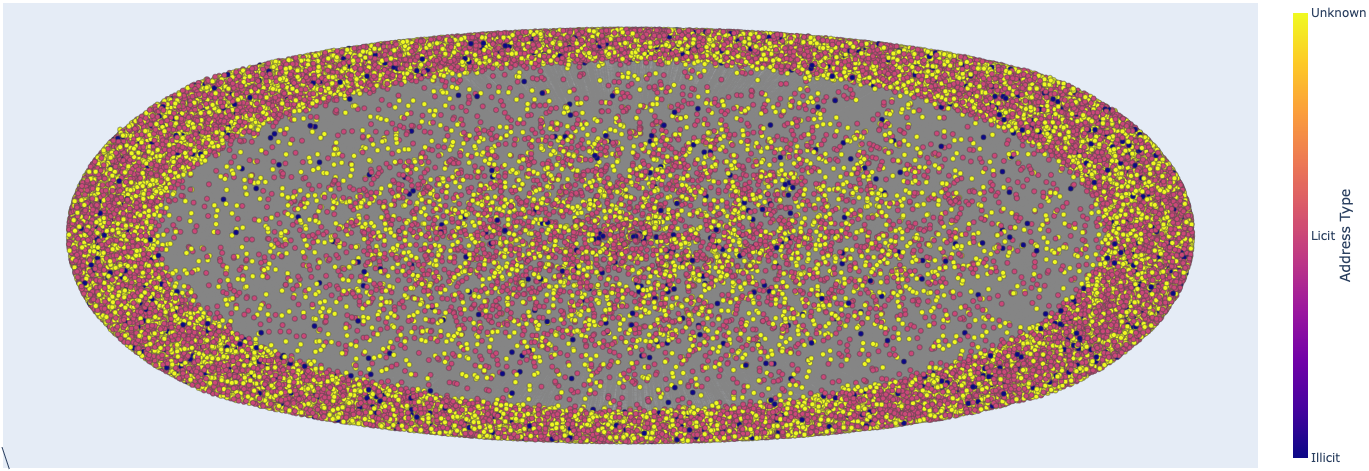}}
}
\vspace{-0.8cm}
\end{figure}

This is a directed graph where nodes represent addresses and edges represent pairwise interactions among input and output addresses of transactions. Figure~\ref{fig:walletstimestep32} visualizes the distribution of all addresses in time step $32$. Unlike Figure~\ref{fig:moneyflowalltxs} where transaction interactions are captured in terms of money flows, Figure~\ref{fig:walletstimestep32} displays the interactions at the address level as opposed to transaction level. The grey area is due to the density of edges. Here, the density of the address graphs at a given time step impact the density of the $k$-hop neighborhood of a wallet address node. We provide a comparison of actor interaction graphs at time steps $14$ and $32$ in the Appendix (see Figure~\ref{fig:walletstimestep1432}). 

\subsubsection{Address-Transaction Graph} \label{addrtxaddr}
\begin{figure}[t]
\vspace{-0.6cm}
\centering{
\caption{An address-transaction graph for selected nodes in time step $32$ featuring Illicit Actor $13$ at the root.}
\hspace*{-0.3cm}\makebox[\linewidth]{\includegraphics[width=3.5in]{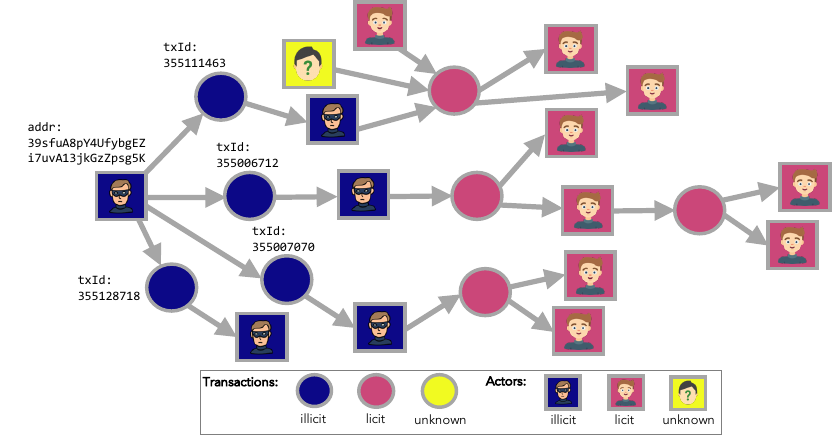}}
\label{fig:addrtxaddrgraph}
}
\vspace{-0.4cm}
\end{figure}

This is a heterogeneous directed graph supported by our Elliptic++ dataset. It has two types of nodes, the transaction node (circle) and the address node (square), and two types of edges, the sender-to-transaction edges, each of which represents the BTC flow from an input address (sender) to a transaction, and the transaction-to-receiver edges, each of which represents the BTC flow from a transaction to an output address. Figure~\ref{fig:addrtxaddrgraph} shows an address-transaction graph of selected transactions and addresses anchored at a given actor (i.e., the Illicit Actor $13$) in time step $32$. The flow and quantity of input and output addresses provide information regarding the purposes of a transaction and the relationships among addresses connected by the same transaction. 
We provide a visual comparison of the distributions of all transactions and addresses at both time steps in the Appendix (see Figure~\ref{fig:txswalletstimestep1432}).

\begin{figure}[b]
\vspace{-0.4cm}
\centering{
\caption{A user entity graph created by grouping transactions with $\geq1$ common element in each set into unique users and adding edges across users if a path existed between groups of transactions in the original graph.}
\vspace{0.1cm}
\makebox[\linewidth]{\includegraphics[width=3in]{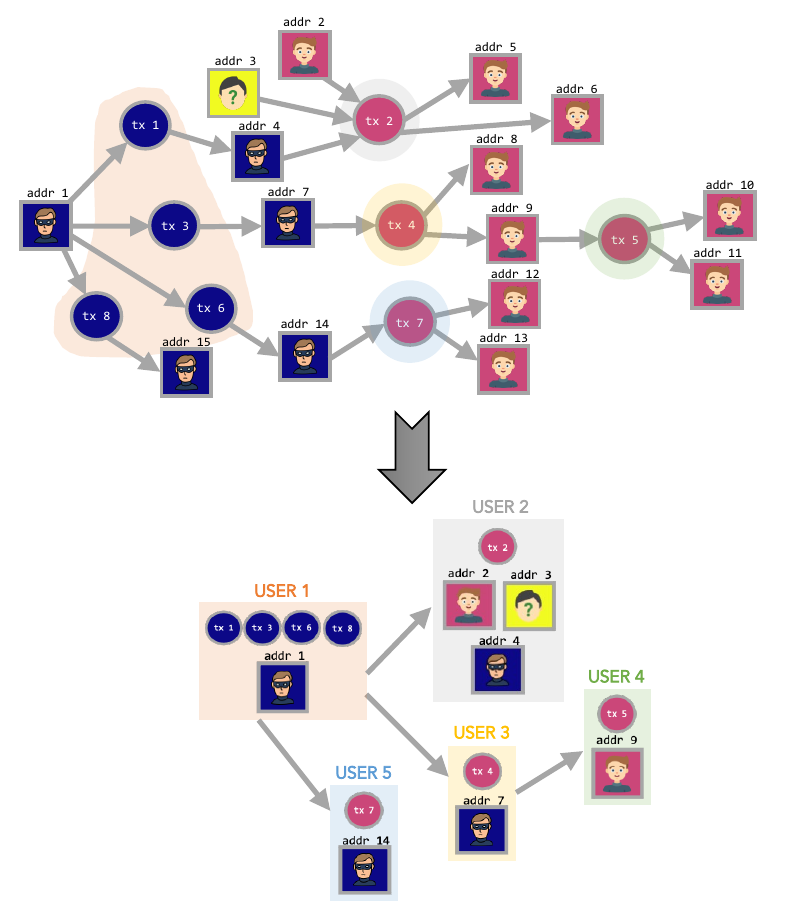}}
\label{fig:fullclusterssmall}
}
\vspace{-0.8cm}
\end{figure} 

\subsubsection{User Entity Graph} \label{userentity}
This graph is generated by address clustering analysis rather than direct construction using the Elliptic++ dataset. 
Clustering addresses involves taking the set of all addresses $\eucal A$ as training data to build a clustering model which creates a set of disjoint subsets $\mathcal U = \left\{ U_1, U_2, ..., U_n \right\}$ (where $U_1, U_2, ..., U_n$ represents $n$ clusters of addresses in $\eucal A$) such that $\bigcup_{i=1}^{n} C_i = \eucal A$. By treating each cluster (sub-graph) as one user, we construct a user-entity graph of $n$ nodes, where each node represents a user in the Bitcoin network and the edges represent interactions between unique users through Blockchain transactions. Address clustering is performed in four steps. 
First, for each transaction, the set of input addresses associated with the transaction is collected. Second, transactions whose address sets are overlapping with one or more addresses are grouped into a unique user. Third, the address-transaction graph is searched to highlight all transactions corresponding to each user. Finally, the highlighted address-transaction graph is converted into a user entity graph. 
Figure~\ref{fig:fullclusterssmall} shows the resulting user graph from the same subset of addresses and transactions in Figure~\ref{fig:addrtxaddrgraph}.
Refer to Section~\ref{clusteringappendix} in the Appendix (Figure~\ref{fig:fullclusters}) for a detailed explanation and visualization of the clustering process. 
Bitcoin address clustering is a hot research topic~\cite{androulaki2013evaluating, reid2013analysis, meiklejohn2013fistful, zhang2020heuristic} due to its potential of linking addresses controlled by a specific user, effectively deanonymizing the identity of those users. 

\begin{table*}
\renewcommand{\arraystretch}{1.3}
\centering 
\caption{The distribution of the number of illicit actors (that appear in $1$, $2-4$, and $\geq 5$ time steps) by time step.}
\label{tab:illicittimesteptable}
\vspace{-0.4cm}
\makebox[\linewidth]{
\begin{tabular}{|c||c|c|c|c|c|c|c|c|c|c|c|c|c|c|c|c|c|}
\hline
\textbf{Time step} & \textbf{1} & \textbf{2} & \textbf{3} & \textbf{4} & \textbf{5} & \textbf{6} & \textbf{7} & \textbf{8} & \textbf{9} & \textbf{10} & \textbf{11} & \textbf{12} & \textbf{13} & \textbf{14} & \textbf{15} & \textbf{16} & \textbf{17}\\
\hline
\hline
illicit actor appears in $1$ time step & 68 & 69 & 42 & 186 & 36 & 8 & 233 & 116 & 384 & 26 & 185 & 47 & 342 & 74 & 193 & 174 & 118\\
\hline
illicit actor appears in $2-4$ time steps & 5 & 1 & 3 & 15 & 2 & 1 & 28 & 7 & 21 & 11 & 7 & 2 & 4 & 0 & 6 & 8 & 8\\ \hline
illicit actor appears in $\geq 5$ time steps & 0 & 0 & 0 & 0 & 2 & 2 & 5 & 4 & 3 & 3 & 2 & 1 & 2 & 0 & 2 & 1 & 6\\ \hline
\end{tabular}
}
\end{table*}
\addtocounter{table}{-1} 
\begin{table*}
\vspace{-0.2cm}
\renewcommand{\arraystretch}{1.3}
\centering
\makebox[\linewidth]{
\begin{tabular}{|c||c|c|c|c|c|c|c|c|c|c|c|c|c|c|c|c|c|}
\hline
\textbf{Time step} & \textbf{18} & \textbf{19} & \textbf{20} & \textbf{21} & \textbf{22} & \textbf{23} & \textbf{24} & \textbf{25} & \textbf{26} & \textbf{27} & \textbf{28} & \textbf{29} & \textbf{30} & \textbf{31} & \textbf{32} &\textbf{33} & \textbf{34}\\
\hline
\hline
illicit actor appears in $1$ time step & 77 & 122 & 405 & 612 & 692 & 376 & 532 & 1096 & 704 & 43 & 97 & 372 & 130 & 131 & 759 & 270 & 447\\
\hline
illicit actor appears in $2-4$ time steps & 4 & 5 & 20 & 16 & 18 & 17 & 40 & 9 & 21 & 1 & 7 & 52 & 11 & 24 & 23 & 7 & 11 \\ \hline
illicit actor appears in $\geq 5$ time steps & 2 & 4 & 5 & 4 & 5 & 3 & 12 & 1 & 9 & 2 & 1 & 9 & 4 & 6 & 8 & 5 & 2\\ \hline
\end{tabular}
}
\end{table*}
\addtocounter{table}{-1} 
\begin{table*}
\vspace{-0.2cm}
\renewcommand{\arraystretch}{1.3}
\centering
\makebox[\linewidth]{
\begin{tabular}{|c||c|c|c|c|c|c|c|c|c|c|c|c|c|c|c|c|c|}
\hline
\textbf{Time step} & \textbf{35} & \textbf{36} & \textbf{37} & \textbf{38} & \textbf{39} & \textbf{40} & \textbf{41} & \textbf{42} & \textbf{43} & \textbf{44} & \textbf{45} & \textbf{46} & \textbf{47} & \textbf{48} & \textbf{49}\\
\hline
\hline
illicit actor appears in $1$ time step & 427 & 588 & 428 & 180 & 127 & 436 & 168 & 395 & 93 & 277 & 23 & 505 & 193 & 370 & 631\\
\hline
illicit actor appears in $2-4$ time steps & 9 & 5 & 5 & 9 & 0 & 14 & 8 & 12 & 11 & 6 & 9 & 4 & 11 & 17 & 20\\ \hline
illicit actor appears in $\geq 5$ time steps & 6 & 4 & 7 & 5 & 2 & 3 & 6 & 3 & 1 & 1 & 1 & 0 & 0 & 1 & 1\\ \hline
\end{tabular}
}
\end{table*}\addtocounter{table}{2} 

\section{Fraud Detection Methodology} \label{methodology}


\vspace{-0.1cm}
\subsection{Dataset Preprocessing}
\begin{figure}
  \vspace{-0.4cm}
  \caption{{\small The distribution of the number of transactions and addresses by time step for training and testing sets.}} \label{fig:distributioncombined}
  \vspace{-0.3cm}
  \makebox[\linewidth]{\subfloat[a][Transactions dataset.]{\includegraphics[width=3.5in]{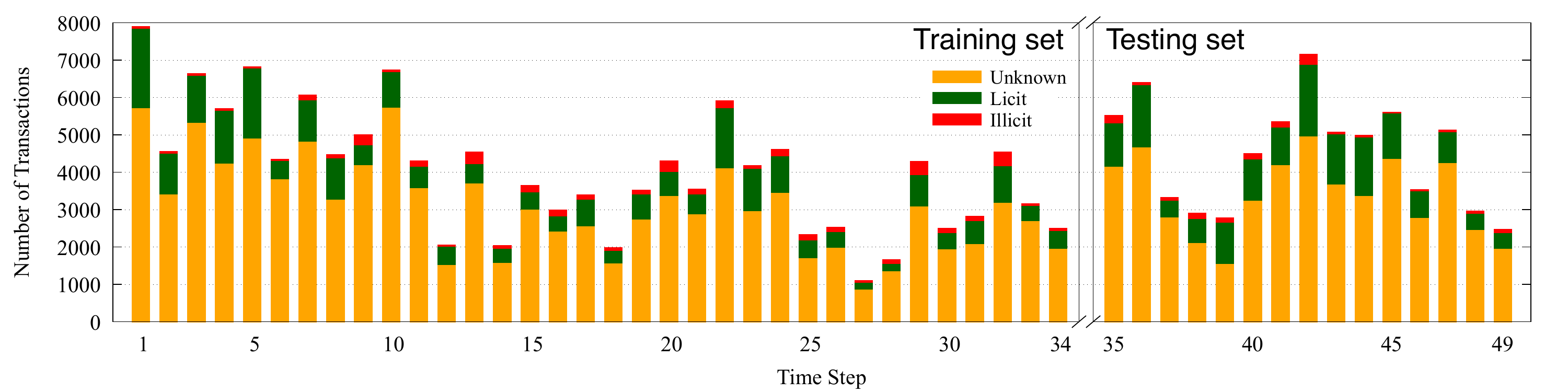}}\label{fig:txspertimestep}}\\
  \vspace{-0.3cm}
  \makebox[\linewidth]{\subfloat[b][Actors dataset.]{\includegraphics[width=3.5in]{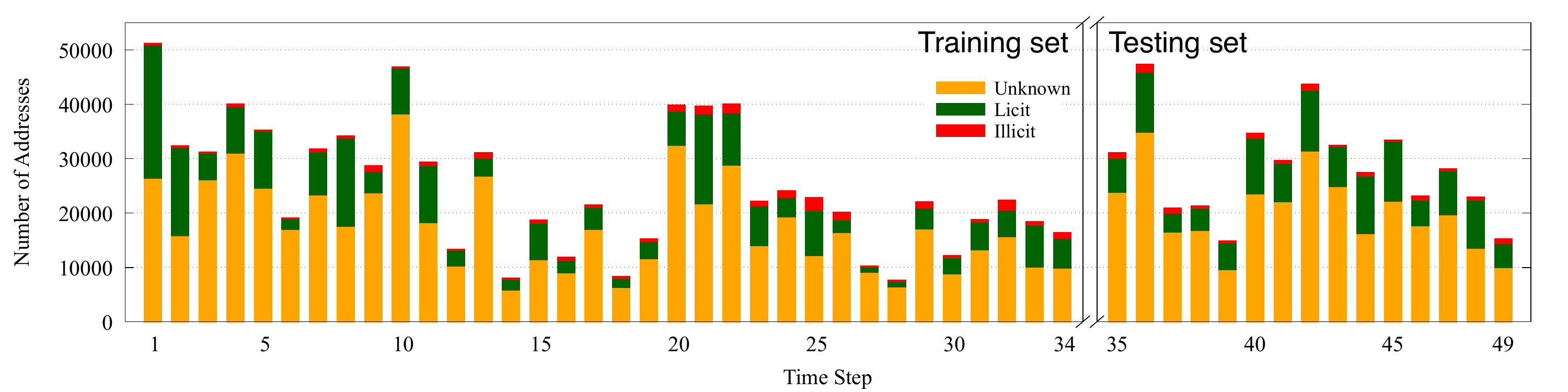}}\label{fig:walletspertimestep}}
  \vspace{-0.6cm}
\end{figure}

We used an effective $70/30$ train-test split with respect to time steps for both the transactions and actors datasets, with time steps $1$ to $34$ for training and time steps $35$ to $49$ for testing. Figure~\ref{fig:distributioncombined} graphically shows the distribution of data points (top for transactions, bottom for actors) of all three classes 
by time step for both the training and testing sets. We provide a detailed distribution of the number of transactions and addresses for all three classes in each of the $49$ steps in the Appendix (Tables~\ref{tab:txspertimesteptable} and~\ref{tab:walletspertimesteptable} respectively). Due to the underlying class imbalance across illicit and licit classes, normalization and standardization transformations are applied. The augmented features in the transactions dataset and all features in the actors dataset are transformed by scaling each feature using the MinMaxScaler to the range $(0,1)$, reducing imbalance and assisting with model convergence~\cite{kramer2016machine}.

\vspace{-0.2cm}
\subsection{Machine Learning Models} \label{classcases}
Previous studies~\cite{oliveira2021guiltywalker,weber2019anti} indicate that ensemble methods perform better than graph neural networks, hence the ML models used for evaluation included Random Forest (RF)~\cite{breiman2001random}, Multilayer Perceptrons (MLP)~\cite{bishop2006pattern}, Long Short-Term Memory (LSTM)~\cite{hochreiter1997long}, and Extreme Gradient Boosting (XGB)~\cite{chen2016xgboost}. We also include Logistic Regression (LR)~\cite{hosmer2013applied} as the baseline. 
LR is a single layer neural network which estimates the probability of an event, such as licit or illicit, based on independent variables. 
RF is an ensemble of classification trees trained on samples with random subsets of features for each decision, evaluating a final decision from averaging all decision trees. 
MLP is an artificial neural network with at least three layers where data features are fed into input neurons that assign probability vectors for classes as outputs. 
The Scikit-learn python library was used for LR (default parameters with 1000 max iterations), RF (default parameters with 50 estimators), and MLP (parameters: 1 hidden layer with 50 neurons, 500 epochs, Adam optimizer, 0.001 learning rate). 
LSTM is a recurrent neural network that has feedback connections and is capable of learning long-term dependencies (sequences of data as compared to single data points). The TensorFlow python library was used for LSTM (hyper-parameters: sigmoid activation, Adam optimizer, 30 epochs, binary cross-entropy loss function, 15 embedding output dims). 
XGB is a supervised learning algorithm which 
predicts variables by combining estimates of prior models. The XGBoost python library was used for XGB (default parameters with "multi:softmax" objective and 2 classes).




\subsection{Fraud Detection Evaluation Metrics}
The metrics used to verify the models were Precision (ratio of correct classifications), Recall (proportion of actual positive labels correctly classified), F1 Score (harmonic mean of precision and recall), and Micro-Avg F1 (Micro-F1) Score (ratio of correct classifications to total classifications). 
In some cases to distinguish between classifiers with close performance, the Matthews Correlation Coefficient (MCC) is used due to its suitability for unbalanced datasets and its prior use on financial fraud and cryptocurrency-related studies~\cite{awoyemi2017credit, zareapoor2015application, zola2019cascading, agarwal2021detecting}. We provide the formal definition of these metrics in Section~\ref{metricsappendix} of the Appendix. In addition, 
to gain deeper understanding of the different ML models and their classification performance for illicit transactions and illicit addresses, we conduct the following three case studies: (i) \texttt{EASY} cases: \textit{all models classify an illicit transaction correctly}; (ii) \texttt{HARD} cases: \textit{all models classify an illicit transaction incorrectly}; and (iii) \texttt{AVERAGE} cases: \textit{some models failed to classify an illicit transaction but $\geq1$ models classified correctly}.

\section{Results and Analysis}

\begin{figure}[t]
\vspace{-0.5cm}
\centering{
\caption{Trend comparison for selected features from the transactions (left) and actors (right) datasets. Green highlight shows good correlation, red shows unclear correlation.}
\label{fig:featurestatsgraphsnew}
\hspace{-0.9cm}\makebox[\linewidth]{\includegraphics[width=4in]{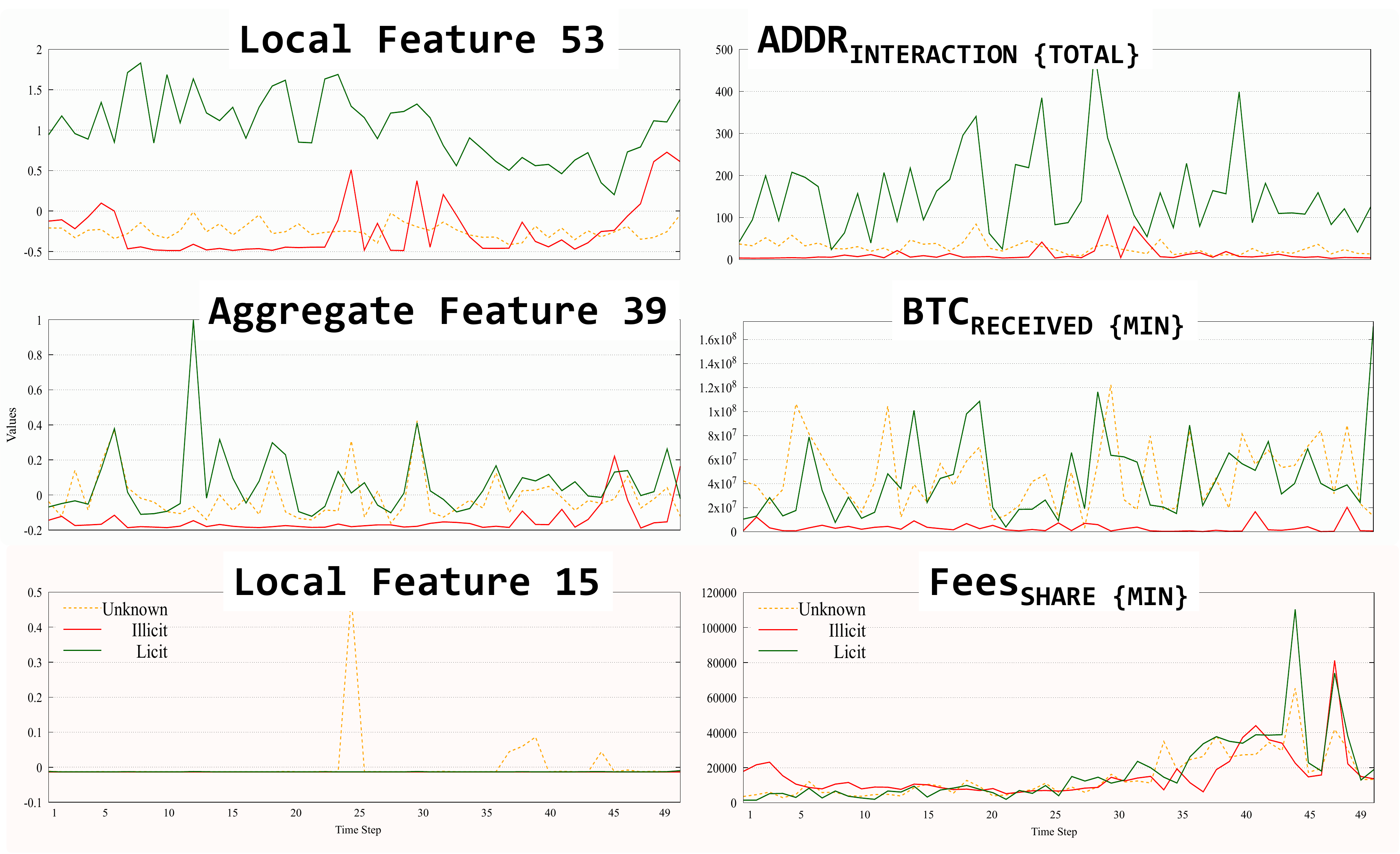}}
}
\vspace{-0.5cm}
\end{figure}

\vspace{-0.1cm}
\subsection{Statistical Analysis of the Dataset} \label{statsanalysis}
The building blocks of the Elliptic++ dataset are the transaction features and address features, which directly impact the quality of information gained by the models and quality of classification explainability into the root cause of fraudulent activities. 
Figure~\ref{fig:featurestatsgraphsnew} shows three chosen features for the transactions (left) and actors (right) datasets, with feature values in y-axis and time steps in x-axis. 
We conjecture that important dataset features will show a reflective trend that clearly distinguishes between the illicit and licit classes. Such features will provide a level of interpretation into the risk assessment of a transaction or an actor. In Figure~\ref{fig:featurestatsgraphsnew}, the green curves of the features in the top two rows show the trend of licit transactions (left) and licit actors (right), which are distinctly different from the red curves of illicit transactions (left) and dishonest actors (right). 
Conversely, some features may not contribute to the detection of illicit transactions and actors, such as the two features in the bottom row. 
For example, the curves for the illicit and licit transactions in Local Feature $53$ can be clearly differentiated through visual analysis, while Local Feature $15$ does not display any visual clue on illicit v.s. licit transactions over all time steps. Similar observations are found in the actors dataset.



\begin{figure}
\vspace{-0.5cm}
\centering{
\caption{Timeline of illicit actors in $\geq 5$ time steps, each point represents at least one illicit transaction involved in.} \vspace{-0.1cm}
\hspace*{0.5cm}\makebox[\linewidth]{\includegraphics[width=4in]{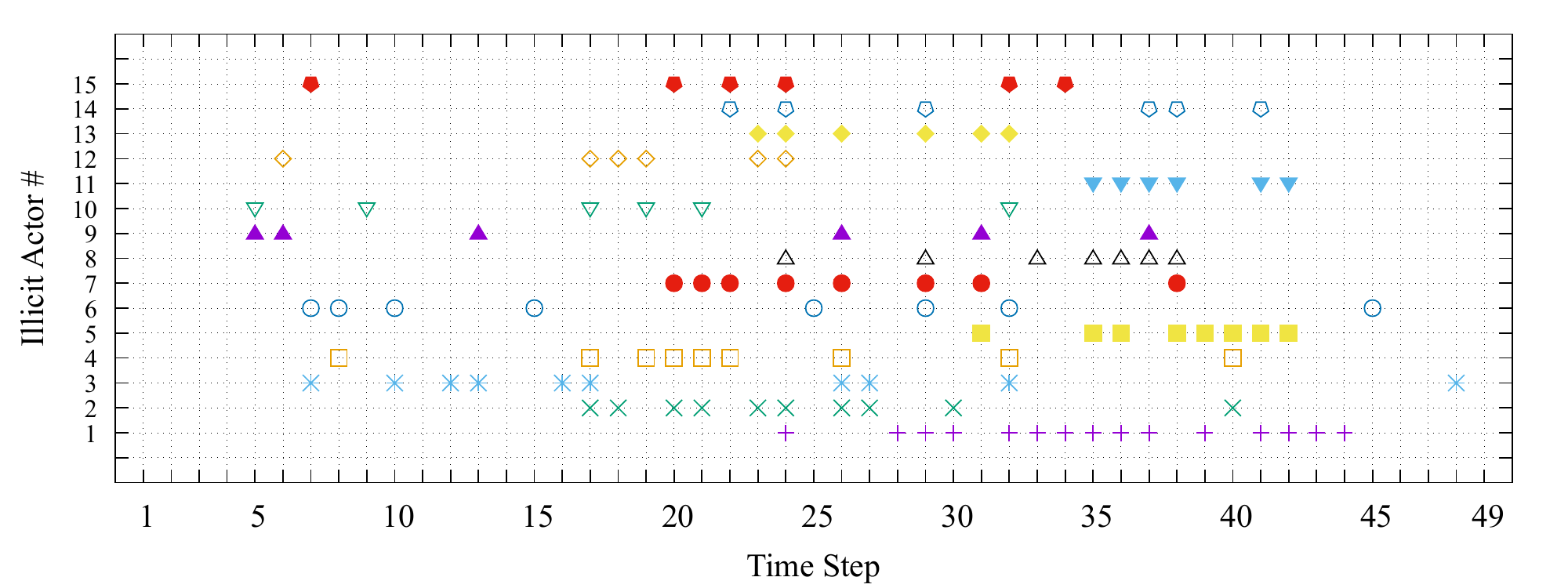}}
\label{fig:badactor}
}
\vspace{-0.5cm}
\end{figure}

\begin{figure}[h]
\vspace{-0.6cm}
\centering{
\caption{Number of transactions per time step for Illicit Actor $13$ (purple), $1$ (green), $5$ (orange), $14$ (Lblue), and $8$ (Dblue).}
\includegraphics[width=3.0in]{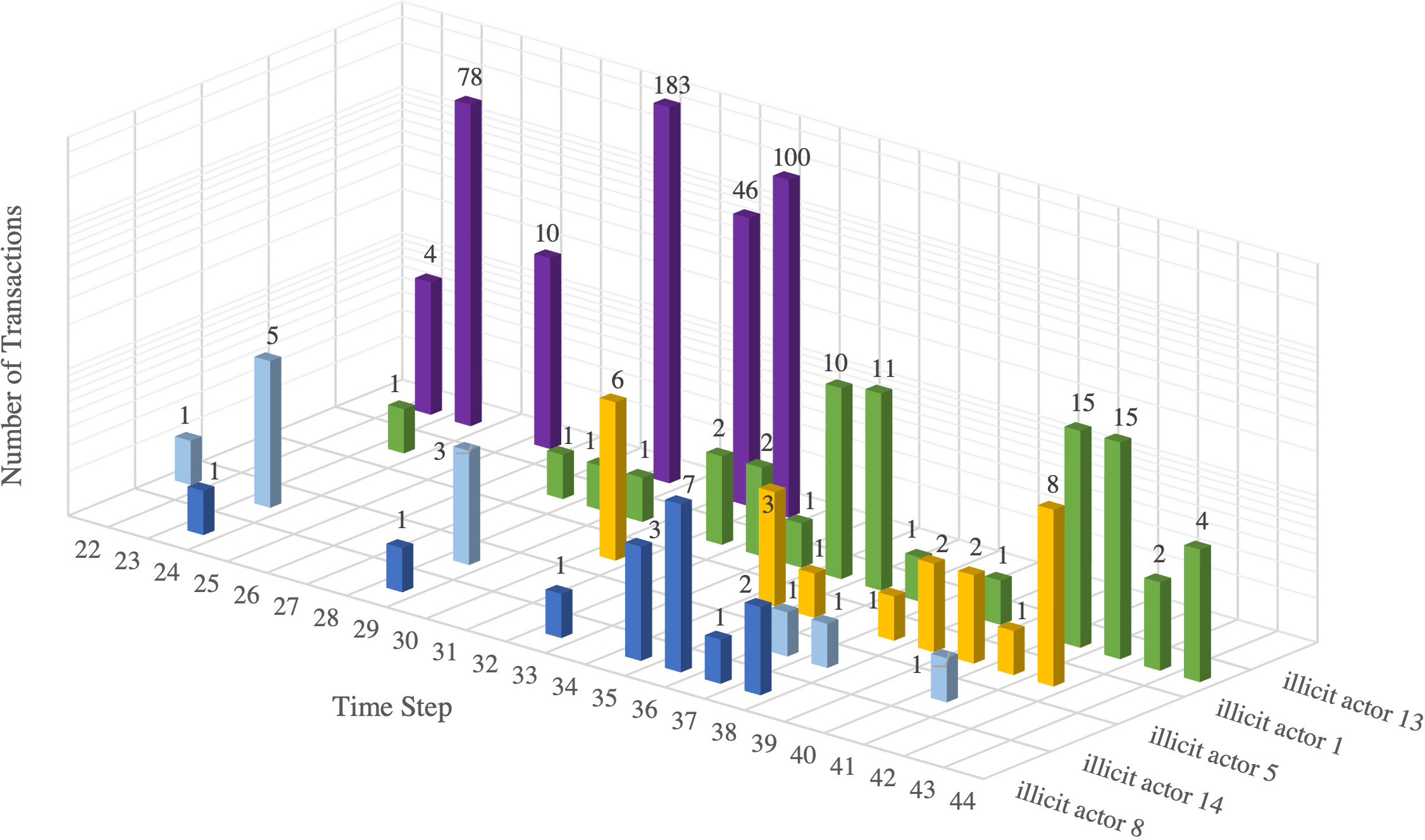}
\label{fig:badactors}
}
\vspace{-0.3cm}
\end{figure}

We can further expand the statistical analysis on the behavioral trends of actors by their features captured in our Elliptic++ dataset, e.g., life span, number of transactions involved, and distribution of actors, which provide additional level of explainability to both the dataset and the classification/detection model performance for illicit/licit/unknown actors. 
Figure~\ref{fig:badactor} shows the timeline of $15$ illicit actors that transact in $\geq 5$ time steps. For instance, Illicit Actor 1 transacts in $15$ time steps, while Illicit Actor 15 transacts in only $6$ time steps. A similar figure for illicit actors existing in only $1$ time step ($14,007$ illicit actors) is included in the Appendix (see Figure~\ref{fig:1000badactor}). Moreover, the number of involved transactions within each time step varies across illicit actors as shown in Figure~\ref{fig:badactors} for the five selected actors. Table~\ref{tab:illicittimesteptable} (shown on the previous page) provides the distribution of illicit actors across time steps categorized into 3 sets.

Regarding the Bitcoin users, clustering addresses using the previously discussed four steps in Section~\ref{userentity}  
created $146,783$ user entities. Table~\ref{tab:statsbitcoinusers} shows some relevant statistics. Each user controlled a varying number of addresses, with $98.72\%$ of users controlling $\leq10$ addresses, while only $0.02\%$ of users controlled $\geq1$K addresses.

\begin{table}[t]
\renewcommand{\arraystretch}{1.3}
\caption{Statistics for Bitcoin users in the Elliptic++ Dataset.}
\label{tab:statsbitcoinusers}
\vspace{-0.2cm}
\centering
\begin{tabular}{|l||c|}
\hline
\textbf{\# Users} & $146,783$\\ \hline
\textbf{\# Addresses per User: Min} & $1$\\ \hline
\textbf{\# Addresses per User: Median} & $1$\\ \hline
\textbf{\# Addresses per User: Mean} & $2.73$\\ \hline
\textbf{\# Addresses per User: Max} & $14,885$\\ \hline
\textbf{\% Users w/ $1-10$ Addresses} & $98.72\%$\\ \hline
\textbf{\% Users w/ $11-1K$ Addresses} & $1.26\%$\\ \hline
\textbf{\% Users w/ $1K-\max$ Addresses} & $0.02\%$\\ \hline
\end{tabular}
\vspace{-0.3cm}
\end{table}

\begin{table}[b]
\renewcommand{\arraystretch}{1.3}
\caption{Illicit transactions results using individual/ensemble of classifiers. \textmd{\texttt{EC}} refers to classification on Elliptic dataset~\cite{weber2019anti}, \textmd{\texttt{TX}} is on our Elliptic++ transactions dataset.}
\label{tab:resultstxs}
\vspace{-0.3cm}
\centering
\begin{tabular}{|c||c|c|c|c|}
\hline
\textbf{Model} & \textbf{Precision} & \textbf{Recall} & \textbf{F1 Score} & \textbf{Micro-F1}\\
\hline
\hline
$\textrm{LR}^{\textrm{EC}}$ & \color[HTML]{FF0000}0.326 & 0.707 & \color[HTML]{FF0000}0.446 & 0.886\\ \hline
$\textrm{LR}^{\textrm{TX}}$ & \color[HTML]{FF0000}0.328 & 0.707 & \color[HTML]{FF0000}0.448 & 0.884\\ \hline
$\textrm{RF}^{\textrm{EC}}$ & 0.940 & 0.724 & 0.818 & 0.979\\ \hline
$\textrm{RF}^{\textrm{TX}}$ & \color[HTML]{008000}\textbf{0.975} & \color[HTML]{008000}\textbf{0.719} & \color[HTML]{008000}\textbf{0.828} & \color[HTML]{008000}\textbf{0.980}\\ \hline
$\textrm{MLP}^{\textrm{EC}}$ & \color[HTML]{FF0000}0.476 & 0.673 & 0.558 & 0.931\\ \hline
$\textrm{MLP}^{\textrm{TX}}$ & 0.611 & 0.613 & 0.612 & 0.949\\ \hline
$\textrm{LSTM}^{\textrm{EC}}$ & 0.665 & \color[HTML]{FF0000}0.350 & \color[HTML]{FF0000}0.459 & 0.946\\ \hline
$\textrm{LSTM}^{\textrm{TX}}$ & 0.709 & \color[HTML]{FF0000}0.223 & \color[HTML]{FF0000}0.339 & 0.942\\ \hline
$\textrm{XGB}^{\textrm{EC}}$ & 0.812 & 0.717 & 0.761 & 0.971\\  \hline 
$\textrm{XGB}^{\textrm{TX}}$ & 0.793 & 0.718 & 0.754 & 0.969\\ \hline\hline
\multicolumn{5}{|c|}{\boldmath{2} \textit{\textbf{classifiers ensemble}, selecting top 3 classifiers}} \\ \hline
$\textrm{RF+MLP}^{\textrm{EC}}$ & 0.987 & 0.624 & 0.765 & 0.975\\ \hline
$\textrm{RF+MLP}^{\textrm{TX}}$ & 0.989 & 0.635 & 0.773 & 0.975\\ \hline
$\textrm{RF+XGB}^{\textrm{EC}}$ & 0.960 & 0.704 & 0.812 & 0.979\\ \hline
$\textrm{RF+XGB}^{\textrm{TX}}$ & \color[HTML]{008000}\textbf{0.977} & \color[HTML]{008000}\textbf{0.706} & \color[HTML]{008000}\textbf{0.820} & \color[HTML]{008000}\textbf{0.979}\\ \hline
$\textrm{MLP+XGB}^{\textrm{EC}}$ & \color[HTML]{FF0000}0.457 & 0.737 & 0.564 & 0.926\\ \hline
$\textrm{MLP+XGB}^{\textrm{TX}}$ & 0.974 & 0.596 & 0.739 & 0.972\\ \hline \hline
\multicolumn{5}{|c|}{\boldmath{3} \textit{\textbf{classifiers ensemble}, selecting top 3 classifiers}} \\ \hline
$\textrm{RF+MLP+XGB}^{\textrm{EC}}$ & 0.947 & 0.719 & 0.817 & 0.979\\ \hline
$\textrm{RF+MLP+XGB}^{\textrm{TX}}$ & \color[HTML]{008000}\textbf{0.962} & \color[HTML]{008000}\textbf{0.723} & \color[HTML]{008000}\textbf{0.826} & \color[HTML]{008000}\textbf{0.980}\\ \hline
\end{tabular}
\end{table}

\begin{table}[b]
\renewcommand{\arraystretch}{1.3}
\caption{Illicit actors results using individual/ensemble of classifiers. \textmd{\texttt{AR}} is classification on our Elliptic++ actors dataset.}
\label{tab:resultsaddr}
\vspace{-0.3cm}
\centering
\begin{tabular}{|c||c|c|c|c|}
\hline
\textbf{Model} & \textbf{Precision} & \textbf{Recall} & \textbf{F1 Score} & \textbf{Micro-F1}\\
\hline
\hline
$\textrm{LR}^{\textrm{AR}}$ & \color[HTML]{FF0000}0.477 & \color[HTML]{FF0000}0.046 & \color[HTML]{FF0000}0.083 & 0.964\\ \hline
$\textrm{RF}^{\textrm{AR}}$ & \color[HTML]{008000}\textbf{0.911} & \color[HTML]{008000}\textbf{0.789} & \color[HTML]{008000}\textbf{0.845} & \color[HTML]{008000}\textbf{0.990}\\ \hline 
$\textrm{MLP}^{\textrm{AR}}$ & 0.708 & 0.502 & 0.587 & 0.974\\ \hline 
$\textrm{LSTM}^{\textrm{AR}}$ & 0.922 & \color[HTML]{FF0000}0.033 & \color[HTML]{FF0000}0.064 & 0.965\\ \hline
$\textrm{XGB}^{\textrm{AR}}$ & 0.869 & 0.534 & 0.662 & 0.980\\ \hline\hline
\multicolumn{5}{|c|}{\boldmath{$2$} \textit{\textbf{classifiers ensemble}, selecting top $3$ classifiers}} \\ \hline
$\textrm{RF+MLP}^{\textrm{AR}}$ & 0.967 & \color[HTML]{FF0000}0.403 & 0.568 & 0.978\\ \hline
$\textrm{RF+XGB}^{\textrm{AR}}$ & \color[HTML]{008000}\textbf{0.959} & \color[HTML]{008000}\textbf{0.530} & \color[HTML]{008000}\textbf{0.682} & \color[HTML]{008000}\textbf{0.982}\\ \hline
$\textrm{MLP+XGB}^{\textrm{AR}}$ & 0.929 & \color[HTML]{FF0000}0.324 & \color[HTML]{FF0000}0.481 & 0.975\\ \hline \hline
\multicolumn{5}{|c|}{\boldmath{$3$} \textit{\textbf{classifiers ensemble}, selecting top $3$ classifiers}} \\ \hline
$\textrm{RF+MLP+XGB}^{\textrm{AR}}$ & \color[HTML]{008000}\textbf{0.933} & \color[HTML]{008000}\textbf{0.572} & \color[HTML]{008000}\textbf{0.709} & \color[HTML]{008000}\textbf{0.983}\\ \hline
\end{tabular}
\end{table}

\vspace{-0.2cm}
\subsection{Model Evaluation and Analysis}

\begin{table*}
\renewcommand{\arraystretch}{1.3}
\centering
\caption{Classification results on the testing dataset showing distributions of \textmd{\texttt{EASY}}, \textmd{\texttt{HARD}}, \textmd{\texttt{AVERAGE}} cases among time steps 35 to 49. Total of each case is $49$, $243$, and $791$ respectively. For the \textmd{\texttt{AVERAGE}} case, distributions are shown for cases where only $1$ model (all shown), only $2$ models (top 2 shown), only $3$ models (top 2 shown), and only $4$ models (top 1 shown) classified correctly.}
\label{tab:resultsbrokendown}
\vspace{-0.3cm}
\begin{tabular}{|cc||c|c|c|c|c|c|c|c|c|c|c|c|c|c|c||c|}
\hline
\multicolumn{2}{|c||}{\textbf{Time Step}} & \textbf{35} & \textbf{36} & \textbf{37} & \textbf{38} & \textbf{39} & \textbf{40} & \textbf{41} & \textbf{42} & \textbf{43} & \textbf{44} & \textbf{45} & \textbf{46} & \textbf{47} & \textbf{48} & \textbf{49} & \textbf{TOTAL}\\ \hline \hline
\multicolumn{2}{|c||}{\texttt{EASY}} & 32 & 0 & 2 & 5 & 5 & 1 & 1 & 3 & 0 & 0 & 0 & 0 & 0 & 0 & 0 & \multicolumn{1}{c|}{49}\\ \hline
\multicolumn{2}{|c||}{\texttt{HARD}} & 4 & 0 & 10 & 7 & 4 & 28 & 6 & 36 & 22 & 20 & 4 & 1 & 21 & 27 & 53 & \multicolumn{1}{c|}{243}\\ \hline
\multicolumn{1}{|c||}{} & LR & 0 & 0 & 3 & 0 & 2 & 3 & 0 & 6 & 2 & 3 & 1 & 0 & 1 & 9 & 2 & \multicolumn{1}{c|}{}\\ \cline{2-17} 
\multicolumn{1}{|c||}{} & RF & 0 & 0 & 1 & 0 & 0 & 0 & 0 & 1 & 0 & 0 & 0 & 0 & 0 & 0 & 0 & \multicolumn{1}{c|}{}\\ \cline{2-17} 
\multicolumn{1}{|c||}{\multirow{6}{*}{\texttt{AVERAGE}}} & MLP & 0 & 0 & 1 & 1 & 0 & 2 & 0 & 2 & 0 & 0 & 0 & 0 & 0 & 0 & 0 & \multicolumn{1}{c|}{\multirow{6}{*}{791}}\\ \cline{2-17} 
\multicolumn{1}{|c||}{} & LSTM & 1 & 0 & 0 & 1 & 0 & 1 & 0 & 0 & 0 & 0 & 0 & 0 & 0 & 0 & 0 & \multicolumn{1}{c|}{}\\ \cline{2-17}
\multicolumn{1}{|c||}{} & XGB & 0 & 0 & 0 & 0 & 0 & 0 & 1 & 2 & 0 & 0 & 0 & 0 & 0 & 0 & 0 & \multicolumn{1}{c|}{}\\ \cline{2-17} 
\multicolumn{1}{|c||}{} & RF,XGB & 4 & 0 & 0 & 1 & 2 & 1 & 17 & 2 & 0 & 0 & 0 & 0 & 0 & 0 & 0 & \multicolumn{1}{c|}{}\\ \cline{2-17} 
\multicolumn{1}{|c||}{} & LR,MLP & 1 & 0 & 0 & 1 & 0 & 2 & 0 & 2 & 0 & 0 & 0 & 0 & 0 & 0 & 0 & \multicolumn{1}{c|}{}\\ \cline{2-17} 
\multicolumn{1}{|c||}{} & RF,MLP,XGB & 5 & 6 & 0 & 8 & 3 & 4 & 1 & 0 & 0 & 0 & 0 & 0 & 0 & 0 & 0 & \multicolumn{1}{c|}{}\\ \cline{2-17} 
\multicolumn{1}{|c||}{} & LR,RF,XGB & 6 & 1 & 10 & 27 & 18 & 10 & 5 & 21 & 0 & 0 & 0 & 0 & 0 & 0 & 0 & \multicolumn{1}{c|}{}\\ \cline{2-17} 
\multicolumn{1}{|c||}{} & RF,MLP,XGB,LR & 124 & 24 & 12 & 57 & 45 & 55 & 81 & 159 & 0 & 1 & 0 & 1 & 0 & 0 & 0 & \multicolumn{1}{c|}{}\\ \hline 
\end{tabular}
\vspace{-0.25cm}
\end{table*}

Table~\ref{tab:resultstxs} and Table~\ref{tab:resultsaddr} show the results of all models trained on the transactions dataset 
and the actors dataset respectively. From Table~\ref{tab:resultstxs}, 
the results for the transactions dataset in Elliptic++ (labelled \texttt{TX}) show an increase in performance in most of the metrics for 
all of the models, compared to the Elliptic dataset (labelled \texttt{EC}). This can be attributed to our addition of $17$ augmented features to each transaction, which in turn improves the generalization performance and the explainability of both the transaction dataset and the fraudulent transaction detection models. 
It is observed that $RF^{TX}$ is the best-performing model (followed by $XGB^{TX}$ and $MLP^{TX}$) with a precision of $97.5\%$ and recall of $71.9\%$. Although $RF^{TX}$ and $RF^{EC}$ have comparable precision and recall performance, $RF^{TX}$ is better as the MCC values are $0.83$ vs $0.81$. 
Table~\ref{tab:resultstxs} also shows the results of $2-$ and $3-$classifier ensembles by selecting the top $3$ models (RF, XGB, MLP). The best performing $2-$ and $3-$classifier ensembles are those using Elliptic++. 
In comparison, the $LR$, $MLP^{EC}$, and $LSTM$ models are ineffective with $<50\%$ precision/recall (highlighted in red). 
Table~\ref{tab:resultsaddr} shows the results for the actors dataset. It is observed that $RF^{AR}$ is the best-performing model (followed by $XGB^{AR}$ and $MLP^{AR}$) with a precision of $91.1\%$ and recall of $78.9\%$. Also, the best $2-$ and $3-$classifier ensembles ($RF+XGB^{AR}$ and $RF+MLP+XGB^{AR}$) show increases in precision ($95.9\%$,$93.3\%$ vs $91.1\%$), but decreases in recall ($53.0\%$,$57.2\%$ vs $78.9\%$), indicating the member models of $2-$ and $3-$classifier ensembles are not complimentary and have high negative correlation~\cite{wu2021boosting}. The $LR^{AR}$ and $LSTM^{AR}$ models are ineffective with extremely low recall (highlighted in red).

\vspace{-0.2cm}
\subsection{\texttt{EASY}, \texttt{HARD}, and \texttt{AVERAGE} cases Analysis}

To further understand the performance of RF models against other models, and the performance results of their best performing  $2-$ and $3-$classifier ensembles, we analyze their performance in terms of the \texttt{EASY}, \texttt{HARD}, and \texttt{AVERAGE} cases as defined in Section 4.3.
Table~\ref{tab:resultsbrokendown} (shown on the next page) provides a temporal split of the classification results across the test time steps of 35 to 49 for the \texttt{EASY} case, \texttt{HARD} case, and \texttt{AVERAGE} case respectively. 
It is important to note that \texttt{AVERAGE} cases present the opportunities for further optimizations.
For the \texttt{AVERAGE} cases (correct classification by $1 \leq x \leq 4$ models), all combinations are shown for each individual model, for the $2/3$ models scenario we show top 2 combinations, and for the $4$ models scenario we show top 1 combination. 
First, the case where the $4$ models RF, MLP, XGB, and LR correctly classify the transaction accounts for $71\%$ of the \texttt{AVERAGE} cases. 
Second, the cases where RF classifies incorrectly only make up
for $<1\%$ of the \texttt{AVERAGE} cases. This motivates us to focus on optimization of the RF model with 
feature refinement.

\begin{figure*}
  \centering
  \vspace{-0.2cm}
  \caption{Top $10$ and bottom $10$ features for the transactions dataset (left) and actors dataset (right).} \label{fig:topbottom}
  \vspace{-0.3cm}
  \makebox[\linewidth]{\subfloat[a][Transactions dataset.]{\includegraphics[width=3.1in]{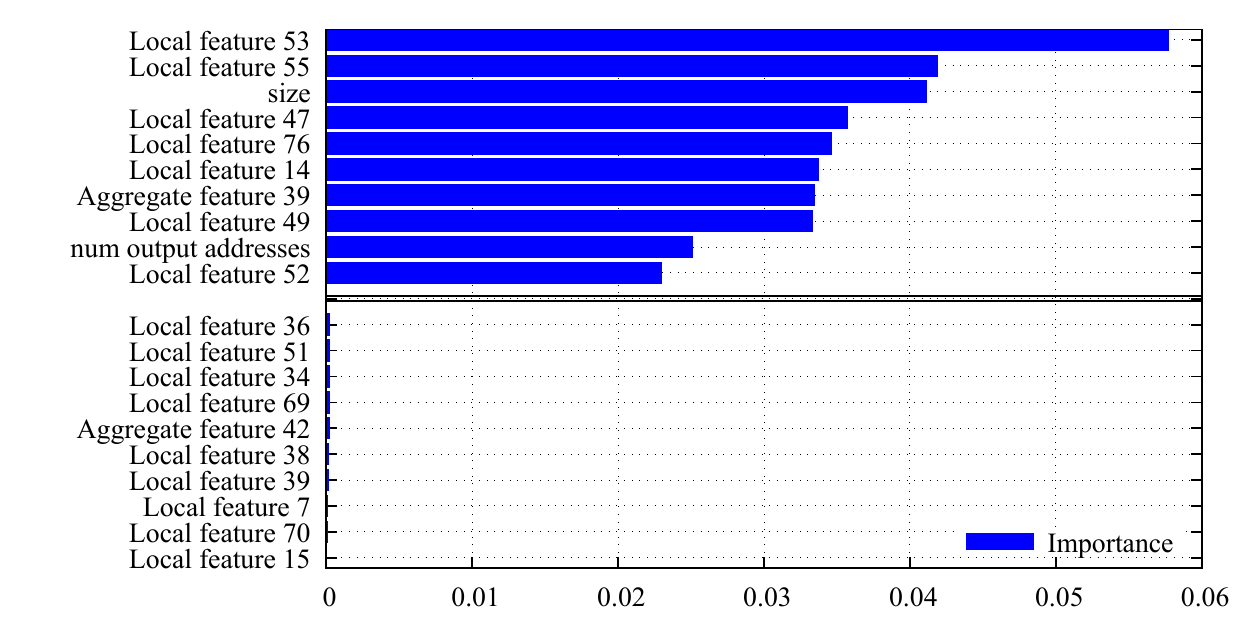}}\label{fig:topbottomtxs}\hspace{2cm}
  \subfloat[b][Actors dataset.]{\includegraphics[width=3.1in]{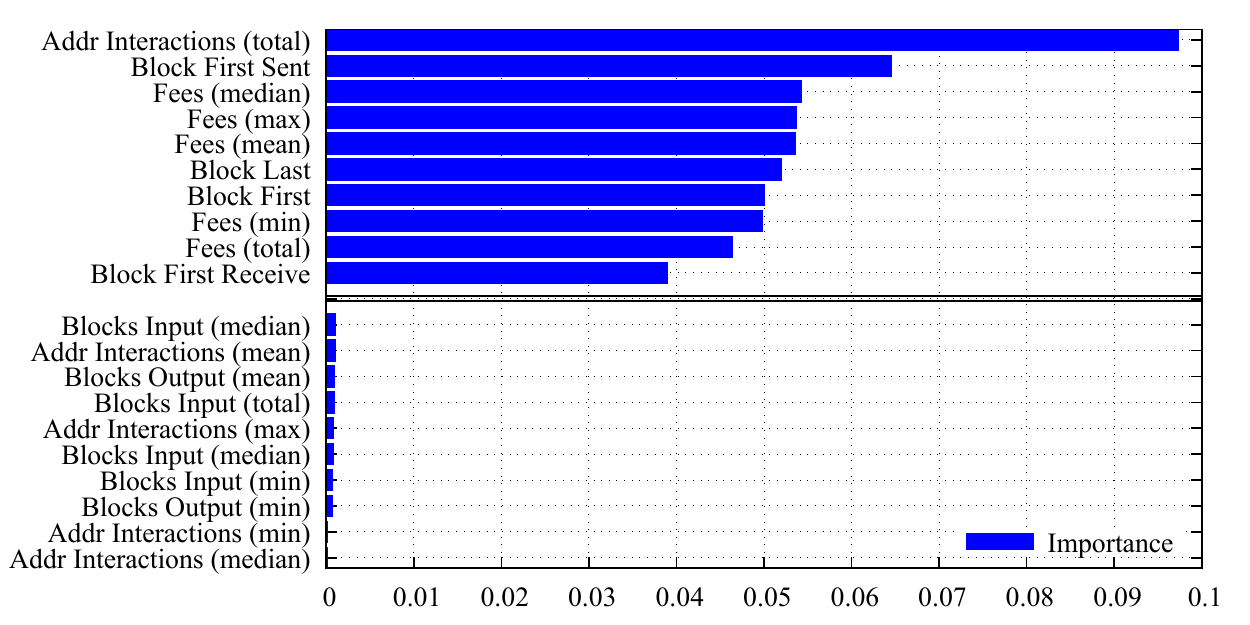}}\label{fig:topbottomaddr}\hspace{0cm}}
  \vspace{-0.2cm}
\end{figure*}

\subsection{Model Optimization by Feature Refinement} \label{featuremodelrefinement}

Given that the RF is the best performing model, we explore feature refinement to further optimize RF. By examining the results from decision trees, we combine feature importance, permutation feature importance, and drop column feature importance to show the top $10$ and bottom $10$ features by importance on both transactions dataset and actors dataset, as shown in Figure~\ref{fig:topbottomtxs}a for \texttt{TX} and Figure~\ref{fig:topbottomaddr}b for \texttt{AR}. For the transactions dataset, the $17$ augmented features produced a collective $12\%$ importance, with the transaction size feature alone responsible for $4.1\%$. For the actors dataset, $35\%$ of the features were responsible for $80\%$ of the importance, with the total address interaction as the most important feature. 
Interestingly, a connection can be made to the set of top and bottom $10$ features with the features highlighted in Figure~\ref{fig:featurestatsgraphsnew}. It can be seen that there is a link from the green highlight to the top $10$ features, and likewise with the red highlight and bottom $10$ features. This solidifies that features providing visual proof of the risk attribute to a larger classification importance, and vice versa. The top and bottom $2$ of each feature type in both datasets are included in the Appendix Section~\ref{statsanalysisappendix} for reference. 
Using this analysis, we run the best $1-$/$2-$/$3-$classifiers ensembles with selected features instead of the full set of features. Tables~\ref{tab:featureoptimizedtransactions} and~\ref{tab:featureoptimizedaddresses} show the model performance results for transactions and wallet addresses (actors) respectively. Interestingly, the feature refined models show an average improvement of $0.92\%$, $1.17\%$, and $0.89\%$ for precision, recall, and F1 score respectively on the transactions dataset, and similarly $1.07\%$, $3.1\%$, and $1.13\%$ on the actors dataset.

\begin{table}[t!]
\renewcommand{\arraystretch}{1.3}
\caption{Illicit transactions classification results using selected features, labelled as $\psi$, 
Micro-F1 denotes Micro-Avg F1.}
\label{tab:featureoptimizedtransactions}
\centering
\vspace{-0.3cm}
\begin{tabular}{|c||c|c|c|c|}
\hline
\textbf{Model} & \textbf{Precision} & \textbf{Recall} & \textbf{F1 Score} & \textbf{Micro-F1}\\
\hline
\hline
$\textrm{RF}^{\textrm{TX}}$ & $0.975$ & $0.719$ & $0.828$ & $0.980$\\ \hline
$\textrm{RF}^{\textrm{TX$^\psi$}}$ & \color[HTML]{008000}$\textbf{0.986}$ & \color[HTML]{008000}$\textbf{0.727}$ & \color[HTML]{008000}$\textbf{0.836}$ & \color[HTML]{008000}$\textbf{0.981}$\\ \hline \hline
$\textrm{RF+XGB}^{\textrm{TX}}$ & $0.977$ & $0.706$ & $0.820$ & $0.979$\\ \hline
$\textrm{RF+XGB}^{\textrm{TX$^\psi$}}$ & \color[HTML]{008000}$\textbf{0.987}$ & \color[HTML]{008000}$\textbf{0.717}$ & \color[HTML]{008000}$\textbf{0.826}$ & \color[HTML]{008000}$\textbf{0.980}$\\ \hline\hline
$\textrm{RF+MLP+XGB}^{\textrm{TX}}$ & $0.962$ & $0.723$ & $0.826$ & $0.980$\\ \hline 
$\textrm{RF+MLP+XGB}^{\textrm{TX$^\psi$}}$ & \color[HTML]{008000}$\textbf{0.968}$ & \color[HTML]{008000}$\textbf{0.729}$ & \color[HTML]{008000}$\textbf{0.834}$ & \color[HTML]{008000}$\textbf{0.980}$\\ \hline 
\end{tabular}
\vspace{-0.9cm}
\end{table}

\begin{figure}[h]
  \centering
  \vspace{-0.6cm}
  \caption{Cumulative votes for 50 RF trees for (a) chosen transactions and (b) addresses before (left) and after (right) feature selection and refinement.} \label{fig:treevotesss}
  \vspace{-0.45cm}
  \makebox[\linewidth]{\subfloat[a][Transactions dataset.]{\includegraphics[width=4.0in]{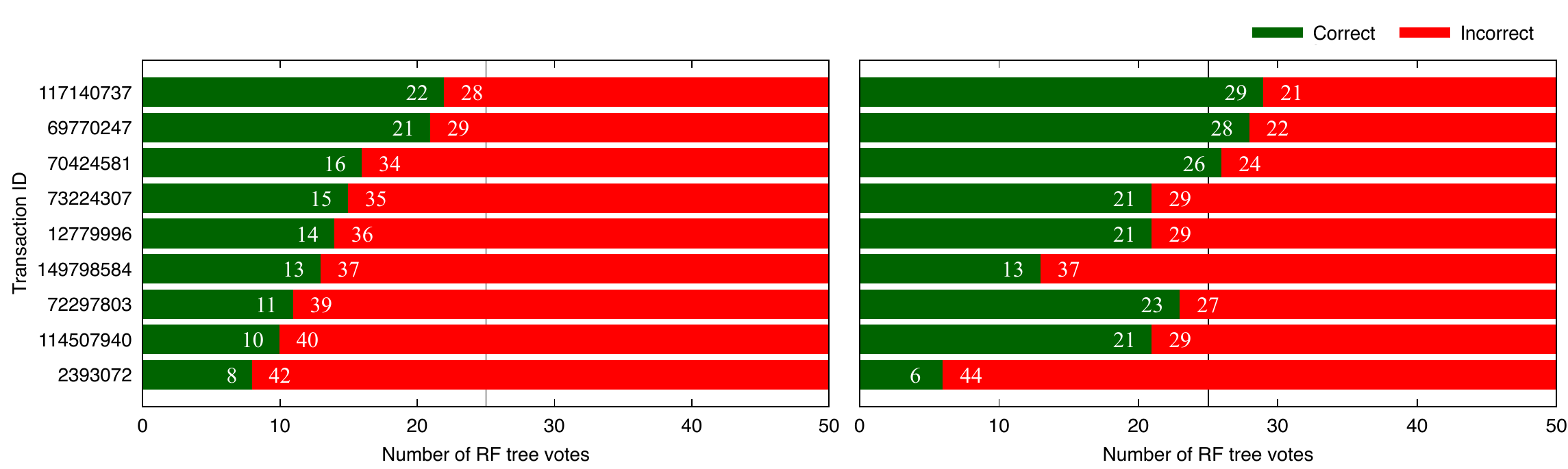}}\label{fig:treevotestxs}\hspace{1.2cm}}\\\vspace{-0.3cm}
  \makebox[\linewidth]{\subfloat[b][Actors dataset.]{\includegraphics[width=4.0in]{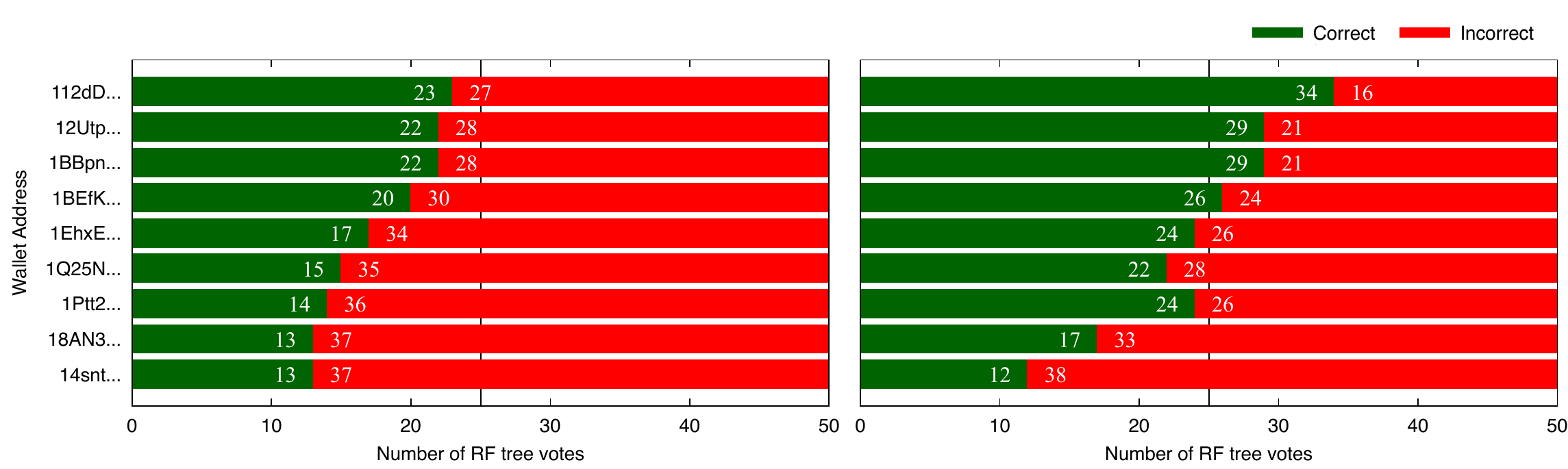}}\label{fig:treevotesaddr}\hspace{1.2cm}}
  \vspace{-0.3cm}
\end{figure}

The increase in recall is also
evident in the RF trees voting. Figures~\ref{fig:treevotestxs}a and~\ref{fig:treevotesaddr}b show the cumulative votes of
$50$ RF trees for $9$ chosen transactions/addresses before (left) and after (right) selecting important features for both the transactions and actors datasets respectively. This demonstrates a trend that when some non-contributing features are dropped, those transactions/addresses that are close to the correct manifold in classification will in turn be classified correctly by more trees,  increasing in the number of correct votes.

\section{Concluding Remarks} \label{conclusion}

With the rapid growth of the cryptocurrency ecosystem, there is a growing demand 
for robust financial forensics on the blockchain networks.
This paper makes three original contributions. 

{\em First}, we collect and contribute the Elliptic++ dataset, combining over 203k transactions and 822k addresses, and providing four graph representations (Money Flow Transaction Graph, Actor Interaction Graph, Address-Transaction Graph, User Entity Graph) that allow for mining and visualizations of connections for anomaly detection. This enables the detection of both illicit transactions and illicit accounts in Bitcoin networks.

{\em Second}, we leverage the four unique graph representations to showcase the fraud detection of illicit transactions, illicit actors (wallet addresses), and the risks of de-anonymization of users using clustering of addresses. We demonstrate the utility of the Elliptic++ dataset for detecting fraudulent transactions and illicit accounts using representative machine learning approaches, including Random Forest, Logistic Regression, Multilayer Perceptrons, LSTM, and Extreme Gradient Boosting. We analyze why Random Forest (RF) is the best-performing model for fraud detection, achieving $97.5\%$ precision and $71.9\%$ recall on the transactions dataset, and $91.1\%$ precision and $78.9\%$ recall on the actors dataset. We also provide comparative analysis of their performance and provide the explainability of the ML algorithms through visualization of all four types of graphs. In addition, we also provide explainability of the performance comparison through detailed analysis on the time series of selected transaction features, and the time series of account features over the time steps of the Elliptic++ dataset.

{\em Third}, to further improve the accuracy of the ML algorithms for detecting fraudulent transactions and illicit accounts, we employ ensemble methods to improve the generalization performance of individual ML algorithms. We study ensemble learning with two and three member models, with detailed analysis on the effectiveness of ensemble methods through \texttt{EASY}, \texttt{HARD} and \texttt{AVERAGE} cases. Motivated by our ensemble learning analysis, we show that model training using selective features instead of all extracted features of transactions and of addresses can further improve RF precision and recall performance by $0.92\%$ and $1.17\%$ respectively for the transactions dataset, and $1.07\%$ and $3.1\%$ respectively for the actors dataset. 
The combination of ensembles and importance-based feature pruning not only demonstrates the utility of the Elliptic++ dataset for fraud detection in Bitcoin network, but also showcases the importance of root cause analysis of fraudulent activities through semantic and statistical explainability of ML algorithms. 

The Bitcoin blockchain is the first and the largest cryptocurrency network, boasting as the most widely used cryptocurrency. We believe that the methodology used for collecting the Elliptic++ dataset and the approach developed to exhibit the utilities of this dataset for fraud detection in Bitcoin network can be leveraged to analyze other types of currencies and blockchain networks, including Monero, ZCash, and Ethereum, to name a few. Moreover, the graph representations and the use of time series representations of transaction/account features over the time steps provides guidelines for developing financial forensics models that can be beneficial to other blockchains with high performance and explainability.

The Elliptic++ dataset and its tutorials are made publicly available at \textbf{\url{https://www.github.com/git-disl/EllipticPlusPlus}}. 
We conjecture that making this dataset publicly available will enable the applied data science researchers to conduct financial forensics on the Bitcoin cryptocurrency network and develop more effective fraud detection models and algorithms. 

\begin{table}
\renewcommand{\arraystretch}{1.3}
\caption{Illicit actors classification results using selected features, labelled as $\psi$ and compared with \textmd{\texttt{AR}}. Micro-F1 denotes Micro-Avg F1.}
\label{tab:featureoptimizedaddresses}
\centering
\vspace{-0.3cm}
\begin{tabular}{|c||c|c|c|c|}
\hline
\textbf{Model} & \textbf{Precision} & \textbf{Recall} & \textbf{F1 Score} & \textbf{Micro-F1}\\
\hline
\hline 
$\textrm{RF}^{\textrm{AR}}$ & 0.911 & 0.789 & 0.845 & 0.990\\ \hline 
$\textrm{RF}^{\textrm{AR$^\psi$}}$ & \color[HTML]{008000}\textbf{0.921} & \color[HTML]{008000}\textbf{0.802} & \color[HTML]{008000}\textbf{0.858} & \color[HTML]{008000}\textbf{0.990}\\ \hline \hline
$\textrm{RF+XGB}^{\textrm{AR}}$ & 0.959 & 0.530 & 0.682 & 0.982\\ \hline
$\textrm{RF+XGB}^{\textrm{AR$^\psi$}}$ & \color[HTML]{008000}\textbf{0.967} & \color[HTML]{008000}\textbf{0.543} & \color[HTML]{008000}\textbf{0.686} & \color[HTML]{008000}\textbf{0.982}\\ \hline \hline
$\textrm{RF+MLP+XGB}^{\textrm{AR}}$ & 0.933 & 0.572 & 0.709 & 0.983\\ \hline 
$\textrm{RF+MLP+XGB}^{\textrm{AR$^\psi$}}$ & \color[HTML]{008000}\textbf{0.945} & \color[HTML]{008000}\textbf{0.601} & \color[HTML]{008000}\textbf{0.718} & \color[HTML]{008000}\textbf{0.984}\\ \hline 
\end{tabular}
\vspace{-0.5cm}
\end{table}

\subsubsection*{Acknowledgement} This research is partially sponsored by the NSF CISE grants 2302720 and 2038029, an IBM faculty award (002114), and a CISCO edge AI grant (001927). Any opinions, findings, and conclusions or recommendations expressed in this material are those of the author(s) and do not necessarily reflect the views of the National Science Foundation or other funding agencies and companies mentioned above.


\bibliographystyle{ACM-Reference-Format}
\balance
\bibliography{main}


\begin{thebibliography}{47}


\ifx \showCODEN    \undefined \def \showCODEN     #1{\unskip}     \fi
\ifx \showDOI      \undefined \def \showDOI       #1{#1}\fi
\ifx \showISBNx    \undefined \def \showISBNx     #1{\unskip}     \fi
\ifx \showISBNxiii \undefined \def \showISBNxiii  #1{\unskip}     \fi
\ifx \showISSN     \undefined \def \showISSN      #1{\unskip}     \fi
\ifx \showLCCN     \undefined \def \showLCCN      #1{\unskip}     \fi
\ifx \shownote     \undefined \def \shownote      #1{#1}          \fi
\ifx \showarticletitle \undefined \def \showarticletitle #1{#1}   \fi
\ifx \showURL      \undefined \def \showURL       {\relax}        \fi
\providecommand\bibfield[2]{#2}
\providecommand\bibinfo[2]{#2}
\providecommand\natexlab[1]{#1}
\providecommand\showeprint[2][]{arXiv:#2}

\bibitem[Agarwal et~al\mbox{.}(2021)]%
        {agarwal2021detecting}
\bibfield{author}{\bibinfo{person}{Rachit Agarwal}, \bibinfo{person}{Shikhar
  Barve}, {and} \bibinfo{person}{Sandeep~Kumar Shukla}.}
  \bibinfo{year}{2021}\natexlab{}.
\newblock \showarticletitle{Detecting malicious accounts in permissionless
  blockchains using temporal graph properties}.
\newblock \bibinfo{journal}{\emph{Applied Network Science}}
  \bibinfo{volume}{6}, \bibinfo{number}{1} (\bibinfo{year}{2021}),
  \bibinfo{pages}{1--30}.
\newblock


\bibitem[Alarab and Prakoonwit(2022)]%
        {alarab2022graph}
\bibfield{author}{\bibinfo{person}{Ismail Alarab} {and} \bibinfo{person}{Simant
  Prakoonwit}.} \bibinfo{year}{2022}\natexlab{}.
\newblock \showarticletitle{Graph-based lstm for anti-money laundering:
  Experimenting temporal graph convolutional network with bitcoin data}.
\newblock \bibinfo{journal}{\emph{Neural Processing Letters}}
  (\bibinfo{year}{2022}), \bibinfo{pages}{1--19}.
\newblock


\bibitem[Alarab et~al\mbox{.}(2020)]%
        {alarab2020comparative}
\bibfield{author}{\bibinfo{person}{Ismail Alarab}, \bibinfo{person}{Simant
  Prakoonwit}, {and} \bibinfo{person}{Mohamed~Ikbal Nacer}.}
  \bibinfo{year}{2020}\natexlab{}.
\newblock \showarticletitle{Comparative analysis using supervised learning
  methods for anti-money laundering in bitcoin}. In
  \bibinfo{booktitle}{\emph{Proceedings of the 2020 5th International
  Conference on Machine Learning Technologies}}. \bibinfo{pages}{11--17}.
\newblock


\bibitem[Androulaki et~al\mbox{.}(2013)]%
        {androulaki2013evaluating}
\bibfield{author}{\bibinfo{person}{Elli Androulaki}, \bibinfo{person}{Ghassan~O
  Karame}, \bibinfo{person}{Marc Roeschlin}, \bibinfo{person}{Tobias Scherer},
  {and} \bibinfo{person}{Srdjan Capkun}.} \bibinfo{year}{2013}\natexlab{}.
\newblock \showarticletitle{Evaluating user privacy in bitcoin}. In
  \bibinfo{booktitle}{\emph{International conference on financial cryptography
  and data security}}. Springer, \bibinfo{pages}{34--51}.
\newblock


\bibitem[Awoyemi et~al\mbox{.}(2017)]%
        {awoyemi2017credit}
\bibfield{author}{\bibinfo{person}{John~O Awoyemi}, \bibinfo{person}{Adebayo~O
  Adetunmbi}, {and} \bibinfo{person}{Samuel~A Oluwadare}.}
  \bibinfo{year}{2017}\natexlab{}.
\newblock \showarticletitle{Credit card fraud detection using machine learning
  techniques: A comparative analysis}. In \bibinfo{booktitle}{\emph{2017
  international conference on computing networking and informatics (ICCNI)}}.
  IEEE, \bibinfo{pages}{1--9}.
\newblock


\bibitem[Bartoletti et~al\mbox{.}(2018)]%
        {bartoletti2018data}
\bibfield{author}{\bibinfo{person}{Massimo Bartoletti},
  \bibinfo{person}{Barbara Pes}, {and} \bibinfo{person}{Sergio Serusi}.}
  \bibinfo{year}{2018}\natexlab{}.
\newblock \showarticletitle{Data mining for detecting bitcoin ponzi schemes}.
  In \bibinfo{booktitle}{\emph{2018 Crypto Valley Conference on Blockchain
  Technology (CVCBT)}}. IEEE, \bibinfo{pages}{75--84}.
\newblock


\bibitem[Biryukov et~al\mbox{.}(2014)]%
        {biryukov2014deanonymisation}
\bibfield{author}{\bibinfo{person}{Alex Biryukov}, \bibinfo{person}{Dmitry
  Khovratovich}, {and} \bibinfo{person}{Ivan Pustogarov}.}
  \bibinfo{year}{2014}\natexlab{}.
\newblock \showarticletitle{Deanonymisation of clients in Bitcoin P2P network}.
  In \bibinfo{booktitle}{\emph{Proceedings of the 2014 ACM SIGSAC Conference on
  Computer and Communications Security}}. \bibinfo{pages}{15--29}.
\newblock


\bibitem[Biryukov and Tikhomirov(2019)]%
        {biryukov2019deanonymization}
\bibfield{author}{\bibinfo{person}{Alex Biryukov} {and} \bibinfo{person}{Sergei
  Tikhomirov}.} \bibinfo{year}{2019}\natexlab{}.
\newblock \showarticletitle{Deanonymization and linkability of cryptocurrency
  transactions based on network analysis}. In \bibinfo{booktitle}{\emph{2019
  IEEE European symposium on security and privacy (EuroS\&P)}}. IEEE,
  \bibinfo{pages}{172--184}.
\newblock


\bibitem[Bishop and Nasrabadi(2006)]%
        {bishop2006pattern}
\bibfield{author}{\bibinfo{person}{Christopher~M Bishop} {and}
  \bibinfo{person}{Nasser~M Nasrabadi}.} \bibinfo{year}{2006}\natexlab{}.
\newblock \bibinfo{booktitle}{\emph{Pattern recognition and machine learning}}.
  Vol.~\bibinfo{volume}{4}.
\newblock \bibinfo{publisher}{Springer}.
\newblock


\bibitem[Breiman(2001)]%
        {breiman2001random}
\bibfield{author}{\bibinfo{person}{Leo Breiman}.}
  \bibinfo{year}{2001}\natexlab{}.
\newblock \showarticletitle{Random forests}.
\newblock \bibinfo{journal}{\emph{Machine learning}} \bibinfo{volume}{45},
  \bibinfo{number}{1} (\bibinfo{year}{2001}), \bibinfo{pages}{5--32}.
\newblock


\bibitem[Chen et~al\mbox{.}(2023)]%
        {chen2023heavy}
\bibfield{author}{\bibinfo{person}{Huiping Chen}, \bibinfo{person}{Grigorios
  Loukides}, \bibinfo{person}{Robert Gwadera}, {and} \bibinfo{person}{Solon~P
  Pissis}.} \bibinfo{year}{2023}\natexlab{}.
\newblock \showarticletitle{Heavy Nodes in a Small Neighborhood: Algorithms and
  Applications}. In \bibinfo{booktitle}{\emph{Proceedings of the 2023 SIAM
  International Conference on Data Mining (SDM)}}. SIAM,
  \bibinfo{pages}{307--315}.
\newblock


\bibitem[Chen and Guestrin(2016)]%
        {chen2016xgboost}
\bibfield{author}{\bibinfo{person}{Tianqi Chen} {and} \bibinfo{person}{Carlos
  Guestrin}.} \bibinfo{year}{2016}\natexlab{}.
\newblock \showarticletitle{Xgboost: A scalable tree boosting system}. In
  \bibinfo{booktitle}{\emph{Proceedings of the 22nd acm sigkdd international
  conference on knowledge discovery and data mining}}.
  \bibinfo{pages}{785--794}.
\newblock


\bibitem[Chen et~al\mbox{.}(2020)]%
        {chen2020phishing}
\bibfield{author}{\bibinfo{person}{Weili Chen}, \bibinfo{person}{Xiongfeng
  Guo}, \bibinfo{person}{Zhiguang Chen}, \bibinfo{person}{Zibin Zheng}, {and}
  \bibinfo{person}{Yutong Lu}.} \bibinfo{year}{2020}\natexlab{}.
\newblock \showarticletitle{Phishing Scam Detection on Ethereum: Towards
  Financial Security for Blockchain Ecosystem.}. In
  \bibinfo{booktitle}{\emph{IJCAI}}. \bibinfo{pages}{4506--4512}.
\newblock


\bibitem[Hirshman et~al\mbox{.}(2013)]%
        {hirshman2013unsupervised}
\bibfield{author}{\bibinfo{person}{Jason Hirshman}, \bibinfo{person}{Yifei
  Huang}, {and} \bibinfo{person}{Stephen Macke}.}
  \bibinfo{year}{2013}\natexlab{}.
\newblock \showarticletitle{Unsupervised approaches to detecting anomalous
  behavior in the bitcoin transaction network}.
\newblock \bibinfo{journal}{\emph{Technical report, Stanford University}}
  (\bibinfo{year}{2013}).
\newblock


\bibitem[Hochreiter and Schmidhuber(1997)]%
        {hochreiter1997long}
\bibfield{author}{\bibinfo{person}{Sepp Hochreiter} {and}
  \bibinfo{person}{J{\"u}rgen Schmidhuber}.} \bibinfo{year}{1997}\natexlab{}.
\newblock \showarticletitle{Long short-term memory}.
\newblock \bibinfo{journal}{\emph{Neural computation}} \bibinfo{volume}{9},
  \bibinfo{number}{8} (\bibinfo{year}{1997}), \bibinfo{pages}{1735--1780}.
\newblock


\bibitem[Hosmer~Jr et~al\mbox{.}(2013)]%
        {hosmer2013applied}
\bibfield{author}{\bibinfo{person}{David~W Hosmer~Jr}, \bibinfo{person}{Stanley
  Lemeshow}, {and} \bibinfo{person}{Rodney~X Sturdivant}.}
  \bibinfo{year}{2013}\natexlab{}.
\newblock \bibinfo{booktitle}{\emph{Applied logistic regression}}.
  Vol.~\bibinfo{volume}{398}.
\newblock \bibinfo{publisher}{John Wiley \& Sons}.
\newblock


\bibitem[Hu et~al\mbox{.}(2019)]%
        {hu2019characterizing}
\bibfield{author}{\bibinfo{person}{Yining Hu}, \bibinfo{person}{Suranga
  Seneviratne}, \bibinfo{person}{Kanchana Thilakarathna},
  \bibinfo{person}{Kensuke Fukuda}, {and} \bibinfo{person}{Aruna Seneviratne}.}
  \bibinfo{year}{2019}\natexlab{}.
\newblock \showarticletitle{Characterizing and detecting money laundering
  activities on the bitcoin network}.
\newblock \bibinfo{journal}{\emph{arXiv preprint arXiv:1912.12060}}
  (\bibinfo{year}{2019}).
\newblock


\bibitem[Kipf and Welling(2016)]%
        {gcn}
\bibfield{author}{\bibinfo{person}{Thomas~N Kipf} {and} \bibinfo{person}{Max
  Welling}.} \bibinfo{year}{2016}\natexlab{}.
\newblock \showarticletitle{Semi-supervised classification with graph
  convolutional networks}.
\newblock \bibinfo{journal}{\emph{arXiv preprint arXiv:1609.02907}}
  (\bibinfo{year}{2016}).
\newblock


\bibitem[Koshy et~al\mbox{.}(2014)]%
        {koshy2014analysis}
\bibfield{author}{\bibinfo{person}{Philip Koshy}, \bibinfo{person}{Diana
  Koshy}, {and} \bibinfo{person}{Patrick McDaniel}.}
  \bibinfo{year}{2014}\natexlab{}.
\newblock \showarticletitle{An analysis of anonymity in bitcoin using p2p
  network traffic}. In \bibinfo{booktitle}{\emph{International Conference on
  Financial Cryptography and Data Security}}. Springer,
  \bibinfo{pages}{469--485}.
\newblock


\bibitem[Kramer(2016)]%
        {kramer2016machine}
\bibfield{author}{\bibinfo{person}{Oliver Kramer}.}
  \bibinfo{year}{2016}\natexlab{}.
\newblock \bibinfo{booktitle}{\emph{Machine learning for evolution
  strategies}}. Vol.~\bibinfo{volume}{20}.
\newblock \bibinfo{publisher}{Springer}.
\newblock


\bibitem[Li et~al\mbox{.}(2022)]%
        {li2022ttagn}
\bibfield{author}{\bibinfo{person}{Sijia Li}, \bibinfo{person}{Gaopeng Gou},
  \bibinfo{person}{Chang Liu}, \bibinfo{person}{Chengshang Hou},
  \bibinfo{person}{Zhenzhen Li}, {and} \bibinfo{person}{Gang Xiong}.}
  \bibinfo{year}{2022}\natexlab{}.
\newblock \showarticletitle{TTAGN: Temporal Transaction Aggregation Graph
  Network for Ethereum Phishing Scams Detection}. In
  \bibinfo{booktitle}{\emph{Proceedings of the ACM Web Conference 2022}}.
  \bibinfo{pages}{661--669}.
\newblock


\bibitem[Lin et~al\mbox{.}(2020a)]%
        {lin2020modeling}
\bibfield{author}{\bibinfo{person}{Dan Lin}, \bibinfo{person}{Jiajing Wu},
  \bibinfo{person}{Qi Yuan}, {and} \bibinfo{person}{Zibin Zheng}.}
  \bibinfo{year}{2020}\natexlab{a}.
\newblock \showarticletitle{Modeling and understanding ethereum transaction
  records via a complex network approach}.
\newblock \bibinfo{journal}{\emph{IEEE Transactions on Circuits and Systems II:
  Express Briefs}} \bibinfo{volume}{67}, \bibinfo{number}{11}
  (\bibinfo{year}{2020}), \bibinfo{pages}{2737--2741}.
\newblock


\bibitem[Lin et~al\mbox{.}(2020b)]%
        {lin2020t}
\bibfield{author}{\bibinfo{person}{Dan Lin}, \bibinfo{person}{Jiajing Wu},
  \bibinfo{person}{Qi Yuan}, {and} \bibinfo{person}{Zibin Zheng}.}
  \bibinfo{year}{2020}\natexlab{b}.
\newblock \showarticletitle{T-edge: Temporal weighted multidigraph embedding
  for ethereum transaction network analysis}.
\newblock \bibinfo{journal}{\emph{Frontiers in Physics}}  \bibinfo{volume}{8}
  (\bibinfo{year}{2020}), \bibinfo{pages}{204}.
\newblock


\bibitem[Loa et~al\mbox{.}(2022)]%
        {loainspection}
\bibfield{author}{\bibinfo{person}{Wai~Weng Loa}, \bibinfo{person}{Gayan~K
  Kulatillekea}, \bibinfo{person}{Mohanad Sarhana}, \bibinfo{person}{Siamak
  Layeghya}, {and} \bibinfo{person}{Marius Portmanna}.}
  \bibinfo{year}{2022}\natexlab{}.
\newblock \showarticletitle{Inspection-L: Self-Supervised GNN Node Embeddings
  for Money Laundering Detection in Bitcoin}.
\newblock  (\bibinfo{year}{2022}).
\newblock


\bibitem[Lorenz et~al\mbox{.}(2020)]%
        {lorenz2020machine}
\bibfield{author}{\bibinfo{person}{Joana Lorenz},
  \bibinfo{person}{Maria~In{\^e}s Silva}, \bibinfo{person}{David
  Apar{\'\i}cio}, \bibinfo{person}{Jo{\~a}o~Tiago Ascens{\~a}o}, {and}
  \bibinfo{person}{Pedro Bizarro}.} \bibinfo{year}{2020}\natexlab{}.
\newblock \showarticletitle{Machine learning methods to detect money laundering
  in the bitcoin blockchain in the presence of label scarcity}. In
  \bibinfo{booktitle}{\emph{Proceedings of the First ACM International
  Conference on AI in Finance}}. \bibinfo{pages}{1--8}.
\newblock


\bibitem[Meiklejohn et~al\mbox{.}(2013)]%
        {meiklejohn2013fistful}
\bibfield{author}{\bibinfo{person}{Sarah Meiklejohn}, \bibinfo{person}{Marjori
  Pomarole}, \bibinfo{person}{Grant Jordan}, \bibinfo{person}{Kirill
  Levchenko}, \bibinfo{person}{Damon McCoy}, \bibinfo{person}{Geoffrey~M
  Voelker}, {and} \bibinfo{person}{Stefan Savage}.}
  \bibinfo{year}{2013}\natexlab{}.
\newblock \showarticletitle{A fistful of bitcoins: characterizing payments
  among men with no names}. In \bibinfo{booktitle}{\emph{Proceedings of the
  2013 conference on Internet measurement conference}}.
  \bibinfo{pages}{127--140}.
\newblock


\bibitem[Monamo et~al\mbox{.}(2016a)]%
        {monamo2016unsupervised}
\bibfield{author}{\bibinfo{person}{Patrick Monamo}, \bibinfo{person}{Vukosi
  Marivate}, {and} \bibinfo{person}{Bheki Twala}.}
  \bibinfo{year}{2016}\natexlab{a}.
\newblock \showarticletitle{Unsupervised learning for robust Bitcoin fraud
  detection}. In \bibinfo{booktitle}{\emph{2016 Information Security for South
  Africa (ISSA)}}. IEEE, \bibinfo{pages}{129--134}.
\newblock


\bibitem[Monamo et~al\mbox{.}(2016b)]%
        {monamo2016multifaceted}
\bibfield{author}{\bibinfo{person}{Patrick~M Monamo}, \bibinfo{person}{Vukosi
  Marivate}, {and} \bibinfo{person}{Bhesipho Twala}.}
  \bibinfo{year}{2016}\natexlab{b}.
\newblock \showarticletitle{A multifaceted approach to Bitcoin fraud detection:
  Global and local outliers}. In \bibinfo{booktitle}{\emph{2016 15th IEEE
  International Conference on Machine Learning and Applications (ICMLA)}}.
  IEEE, \bibinfo{pages}{188--194}.
\newblock


\bibitem[Nakamoto(2008)]%
        {nakamoto2008bitcoin}
\bibfield{author}{\bibinfo{person}{Satoshi Nakamoto}.}
  \bibinfo{year}{2008}\natexlab{}.
\newblock \showarticletitle{Bitcoin: A peer-to-peer electronic cash system}.
\newblock \bibinfo{journal}{\emph{Decentralized Business Review}}
  (\bibinfo{year}{2008}), \bibinfo{pages}{21260}.
\newblock


\bibitem[Oliveira et~al\mbox{.}(2021)]%
        {oliveira2021guiltywalker}
\bibfield{author}{\bibinfo{person}{Catarina Oliveira},
  \bibinfo{person}{Jo{\~a}o Torres}, \bibinfo{person}{Maria~In{\^e}s Silva},
  \bibinfo{person}{David Apar{\'\i}cio}, \bibinfo{person}{Jo{\~a}o~Tiago
  Ascens{\~a}o}, {and} \bibinfo{person}{Pedro Bizarro}.}
  \bibinfo{year}{2021}\natexlab{}.
\newblock \showarticletitle{GuiltyWalker: Distance to illicit nodes in the
  Bitcoin network}.
\newblock \bibinfo{journal}{\emph{arXiv preprint arXiv:2102.05373}}
  (\bibinfo{year}{2021}).
\newblock


\bibitem[Pham and Lee(2016a)]%
        {pham2016anomaly}
\bibfield{author}{\bibinfo{person}{Thai Pham} {and} \bibinfo{person}{Steven
  Lee}.} \bibinfo{year}{2016}\natexlab{a}.
\newblock \showarticletitle{Anomaly detection in bitcoin network using
  unsupervised learning methods}.
\newblock \bibinfo{journal}{\emph{arXiv preprint arXiv:1611.03941}}
  (\bibinfo{year}{2016}).
\newblock


\bibitem[Pham and Lee(2016b)]%
        {pham2016anomaly2}
\bibfield{author}{\bibinfo{person}{Thai Pham} {and} \bibinfo{person}{Steven
  Lee}.} \bibinfo{year}{2016}\natexlab{b}.
\newblock \showarticletitle{Anomaly detection in the bitcoin system-a network
  perspective}.
\newblock \bibinfo{journal}{\emph{arXiv preprint arXiv:1611.03942}}
  (\bibinfo{year}{2016}).
\newblock


\bibitem[R.~Michalski and Macek(2020)]%
        {michalski+2020-IEEEaccess}
\bibfield{author}{\bibinfo{person}{D.~Dziubaltowska R.~Michalski} {and}
  \bibinfo{person}{P. Macek}.} \bibinfo{year}{2020}\natexlab{}.
\newblock \showarticletitle{Revealing the character of nodes in a blockchain
  with supervised learning}.
\newblock \bibinfo{journal}{\emph{IEEE Access}} (\bibinfo{year}{2020}).
\newblock


\bibitem[Reid and Harrigan(2013)]%
        {reid2013analysis}
\bibfield{author}{\bibinfo{person}{Fergal Reid} {and} \bibinfo{person}{Martin
  Harrigan}.} \bibinfo{year}{2013}\natexlab{}.
\newblock \showarticletitle{An analysis of anonymity in the bitcoin system}.
\newblock In \bibinfo{booktitle}{\emph{Security and privacy in social
  networks}}. \bibinfo{publisher}{Springer}, \bibinfo{pages}{197--223}.
\newblock


\bibitem[Shen et~al\mbox{.}(2021)]%
        {shen2021identity}
\bibfield{author}{\bibinfo{person}{Jie Shen}, \bibinfo{person}{Jiajun Zhou},
  \bibinfo{person}{Yunyi Xie}, \bibinfo{person}{Shanqing Yu}, {and}
  \bibinfo{person}{Qi Xuan}.} \bibinfo{year}{2021}\natexlab{}.
\newblock \showarticletitle{Identity inference on blockchain using graph neural
  network}. In \bibinfo{booktitle}{\emph{International Conference on Blockchain
  and Trustworthy Systems}}. Springer, \bibinfo{pages}{3--17}.
\newblock


\bibitem[Sihao~Hu and Liu(2023a)]%
        {Sihao+WWW2023}
\bibfield{author}{\bibinfo{person}{Bingqiao Luo Shengliang Lu Bingsheng~He
  Sihao~Hu, Zhen~Zhang} {and} \bibinfo{person}{Ling Liu}.}
  \bibinfo{year}{2023}\natexlab{a}.
\newblock \showarticletitle{BERT4ETH: Pre-training of Transformer Encoder
  Representation for Ethereum Fraud Detection}. In
  \bibinfo{booktitle}{\emph{The Web Conference}}. May 2-4, 2023, Austin, TX).
\newblock


\bibitem[Sihao~Hu and Liu(2023b)]%
        {Sihao+PSBert-2023}
\bibfield{author}{\bibinfo{person}{Ka~Ho Chow Fatih Ilhan Selim~Tekin Sihao~Hu,
  Tiansheg~Huang} {and} \bibinfo{person}{Ling Liu}.}
  \bibinfo{year}{2023}\natexlab{b}.
\newblock \showarticletitle{Etherum Account Profiling and De-anonymization via
  Pseudo-Siamese BERT}.
\newblock \bibinfo{journal}{\emph{Technical Report, Georgia Tech}}
  (\bibinfo{year}{2023}).
\newblock


\bibitem[Sun et~al\mbox{.}(2022)]%
        {sun2022monlad}
\bibfield{author}{\bibinfo{person}{Xiaobing Sun}, \bibinfo{person}{Wenjie
  Feng}, \bibinfo{person}{Shenghua Liu}, \bibinfo{person}{Yuyang Xie},
  \bibinfo{person}{Siddharth Bhatia}, \bibinfo{person}{Bryan Hooi},
  \bibinfo{person}{Wenhan Wang}, {and} \bibinfo{person}{Xueqi Cheng}.}
  \bibinfo{year}{2022}\natexlab{}.
\newblock \showarticletitle{MonLAD: Money laundering agents detection in
  transaction streams}. In \bibinfo{booktitle}{\emph{Proceedings of the
  Fifteenth ACM International Conference on Web Search and Data Mining}}.
  \bibinfo{pages}{976--986}.
\newblock


\bibitem[Veli{\v{c}}kovi{\'c} et~al\mbox{.}(2017)]%
        {gat}
\bibfield{author}{\bibinfo{person}{Petar Veli{\v{c}}kovi{\'c}},
  \bibinfo{person}{Guillem Cucurull}, \bibinfo{person}{Arantxa Casanova},
  \bibinfo{person}{Adriana Romero}, \bibinfo{person}{Pietro Lio}, {and}
  \bibinfo{person}{Yoshua Bengio}.} \bibinfo{year}{2017}\natexlab{}.
\newblock \showarticletitle{Graph attention networks}.
\newblock \bibinfo{journal}{\emph{arXiv preprint arXiv:1710.10903}}
  (\bibinfo{year}{2017}).
\newblock


\bibitem[Weber et~al\mbox{.}(2019)]%
        {weber2019anti}
\bibfield{author}{\bibinfo{person}{Mark Weber}, \bibinfo{person}{Giacomo
  Domeniconi}, \bibinfo{person}{Jie Chen}, \bibinfo{person}{Daniel Karl~I
  Weidele}, \bibinfo{person}{Claudio Bellei}, \bibinfo{person}{Tom Robinson},
  {and} \bibinfo{person}{Charles~E Leiserson}.}
  \bibinfo{year}{2019}\natexlab{}.
\newblock \showarticletitle{Anti-money laundering in bitcoin: Experimenting
  with graph convolutional networks for financial forensics}.
\newblock \bibinfo{journal}{\emph{arXiv preprint arXiv:1908.02591}}
  (\bibinfo{year}{2019}).
\newblock


\bibitem[Wu et~al\mbox{.}(2021a)]%
        {wu2021detecting}
\bibfield{author}{\bibinfo{person}{Jiajing Wu}, \bibinfo{person}{Jieli Liu},
  \bibinfo{person}{Weili Chen}, \bibinfo{person}{Huawei Huang},
  \bibinfo{person}{Zibin Zheng}, {and} \bibinfo{person}{Yan Zhang}.}
  \bibinfo{year}{2021}\natexlab{a}.
\newblock \showarticletitle{Detecting mixing services via mining bitcoin
  transaction network with hybrid motifs}.
\newblock \bibinfo{journal}{\emph{IEEE Transactions on Systems, Man, and
  Cybernetics: Systems}} \bibinfo{volume}{52}, \bibinfo{number}{4}
  (\bibinfo{year}{2021}), \bibinfo{pages}{2237--2249}.
\newblock


\bibitem[Wu et~al\mbox{.}(2020)]%
        {trans2vec}
\bibfield{author}{\bibinfo{person}{Jiajing Wu}, \bibinfo{person}{Qi Yuan},
  \bibinfo{person}{Dan Lin}, \bibinfo{person}{Wei You}, \bibinfo{person}{Weili
  Chen}, \bibinfo{person}{Chuan Chen}, {and} \bibinfo{person}{Zibin Zheng}.}
  \bibinfo{year}{2020}\natexlab{}.
\newblock \showarticletitle{Who are the phishers? phishing scam detection on
  ethereum via network embedding}.
\newblock \bibinfo{journal}{\emph{IEEE Transactions on Systems, Man, and
  Cybernetics: Systems}} (\bibinfo{year}{2020}).
\newblock


\bibitem[Wu et~al\mbox{.}(2021b)]%
        {wu2021boosting}
\bibfield{author}{\bibinfo{person}{Yanzhao Wu}, \bibinfo{person}{Ling Liu},
  \bibinfo{person}{Zhongwei Xie}, \bibinfo{person}{Ka-Ho Chow}, {and}
  \bibinfo{person}{Wenqi Wei}.} \bibinfo{year}{2021}\natexlab{b}.
\newblock \showarticletitle{Boosting ensemble accuracy by revisiting ensemble
  diversity metrics}. In \bibinfo{booktitle}{\emph{Proceedings of the IEEE/CVF
  Conference on Computer Vision and Pattern Recognition}}.
  \bibinfo{pages}{16469--16477}.
\newblock


\bibitem[Zareapoor et~al\mbox{.}(2015)]%
        {zareapoor2015application}
\bibfield{author}{\bibinfo{person}{Masoumeh Zareapoor}, \bibinfo{person}{Pourya
  Shamsolmoali}, {et~al\mbox{.}}} \bibinfo{year}{2015}\natexlab{}.
\newblock \showarticletitle{Application of credit card fraud detection: Based
  on bagging ensemble classifier}.
\newblock \bibinfo{journal}{\emph{Procedia computer science}}
  \bibinfo{volume}{48}, \bibinfo{number}{2015} (\bibinfo{year}{2015}),
  \bibinfo{pages}{679--685}.
\newblock


\bibitem[Zhang et~al\mbox{.}(2020)]%
        {zhang2020heuristic}
\bibfield{author}{\bibinfo{person}{Yuhang Zhang}, \bibinfo{person}{Jun Wang},
  {and} \bibinfo{person}{Jie Luo}.} \bibinfo{year}{2020}\natexlab{}.
\newblock \showarticletitle{Heuristic-based address clustering in bitcoin}.
\newblock \bibinfo{journal}{\emph{IEEE Access}}  \bibinfo{volume}{8}
  (\bibinfo{year}{2020}), \bibinfo{pages}{210582--210591}.
\newblock


\bibitem[Zhou et~al\mbox{.}(2022)]%
        {zhou2022behavior}
\bibfield{author}{\bibinfo{person}{Jiajun Zhou}, \bibinfo{person}{Chenkai Hu},
  \bibinfo{person}{Jianlei Chi}, \bibinfo{person}{Jiajing Wu},
  \bibinfo{person}{Meng Shen}, {and} \bibinfo{person}{Qi Xuan}.}
  \bibinfo{year}{2022}\natexlab{}.
\newblock \showarticletitle{Behavior-aware Account De-anonymization on Ethereum
  Interaction Graph}.
\newblock \bibinfo{journal}{\emph{arXiv preprint arXiv:2203.09360}}
  (\bibinfo{year}{2022}).
\newblock


\bibitem[Zola et~al\mbox{.}(2019)]%
        {zola2019cascading}
\bibfield{author}{\bibinfo{person}{Francesco Zola}, \bibinfo{person}{Maria
  Eguimendia}, \bibinfo{person}{Jan~Lukas Bruse}, {and}
  \bibinfo{person}{Raul~Orduna Urrutia}.} \bibinfo{year}{2019}\natexlab{}.
\newblock \showarticletitle{Cascading machine learning to attack bitcoin
  anonymity}. In \bibinfo{booktitle}{\emph{2019 IEEE International Conference
  on Blockchain (Blockchain)}}. IEEE, \bibinfo{pages}{10--17}.
\newblock


\end{thebibliography}

\clearpage
\appendix

\section{Appendix}

\subsection{Dataset Collection}
\begin{figure*}
\centering{
\caption{Elliptic++ dataset collection pipeline using our blockchain parsers and extractors.}
\makebox[\linewidth]{\includegraphics[width=8in]{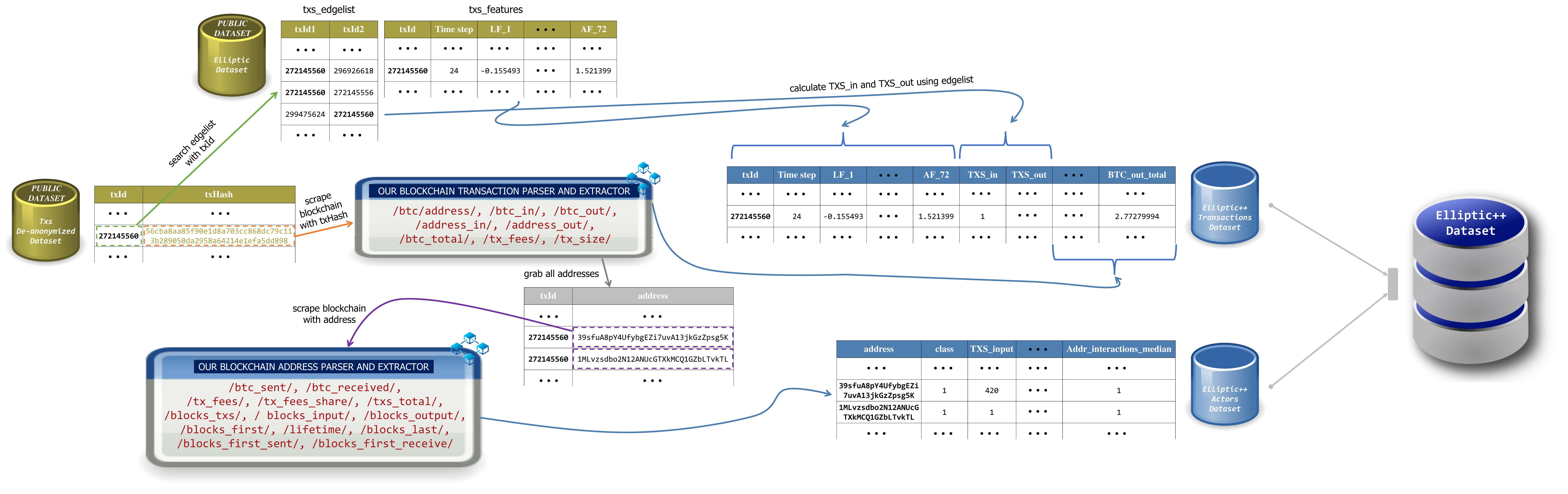}}
\label{fig:datasetcreation}
}
\end{figure*}

The Bitcoin blockchain contains a huge number of transactions since its inception. Given that our goal of data collection is for fraud detection on the Bitcoin network, we use the 203k transactions collected in the Elliptic dataset as the seeds of our crawling module for two reasons. First, the Elliptic dataset is the largest labelled Bitcoin transaction data publicly available. Second, the Elliptic dataset only contains transactions in the Bitcoin network labeled as licit, illicit, and unknown. To enable detection of both fraudulent transactions and illicit accounts, we need to collect all accounts involved in the Elliptic transaction dataset. The data acquisition pipeline for the Elliptic++ dataset is shown in Figure~\ref{fig:datasetcreation} and is described as follows.

First, we design and implement a blockchain parser which takes de-anonymized transactions as input, scrapes the public Blockchain, and augments additional features to the transactions. The augmented features include information regarding the transaction including number of incoming/outgoing edges, number of input/output addresses, amount of incoming/outgoing BTC, and much more

Next, the output from the blockchain transactions parser is fed into our extractor that parses the addresses involved in the transactions and scrapes the public Blockchain to gather the corresponding address-related features. These features include transaction and time related information.

We define the transaction data structure (see Table~\ref{tab:onetxsexample}) and account data structure (see Table~\ref{tab:oneaddrexample}) such that the parsers and extractors will organize, clean, and store the data based on these structures. Furthermore, we collect those account features that are needed in the construction of the four types of graph representations: (a) Money Flow Transaction Graph, (b) Actor Interaction Graph, (c) Address-Transaction Graph, and (d) User Entity Graph. These graphs enable mining and visualizations of account and transaction connections for anomaly detection.

In addition, to identify and define the important features for all accounts involved in the Elliptic transaction dataset, as outlined above, a major technical challenge was the resource demand. The collection, cleaning, parsing, and labeling of the dataset were computation and storage intensive. We leveraged the High Performance Computing (HPC) Clusters at the Georgia Institute of Technology (the PACE clusters\footnote{\url{www.pace.gatech.edu}}), taking several weeks to complete the extraction of all accounts and their features present in the Elliptic++ dataset.

\begin{figure*}
\centering{
\caption{Distribution of transactions in time steps $14$ (left) and $32$ (right).}\vspace{0.3cm}
\makebox[\linewidth]{\includegraphics[width=7in]{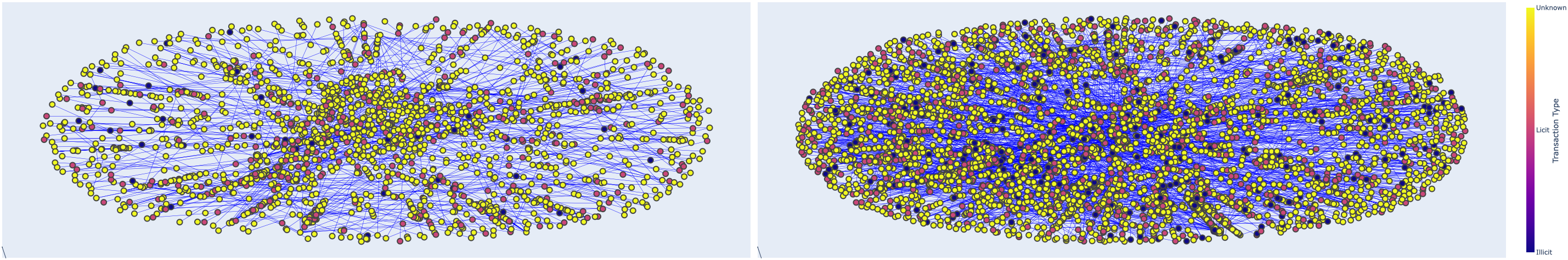}}
\label{fig:txstimestep1432}
}
\vspace{-0.7cm}
\end{figure*} 

\begin{figure*}
\centering{
\caption{Distribution of wallet addresses in time steps $14$ (left) and $32$ (right).}\vspace{0.3cm}
\makebox[\linewidth]{\includegraphics[width=7in]{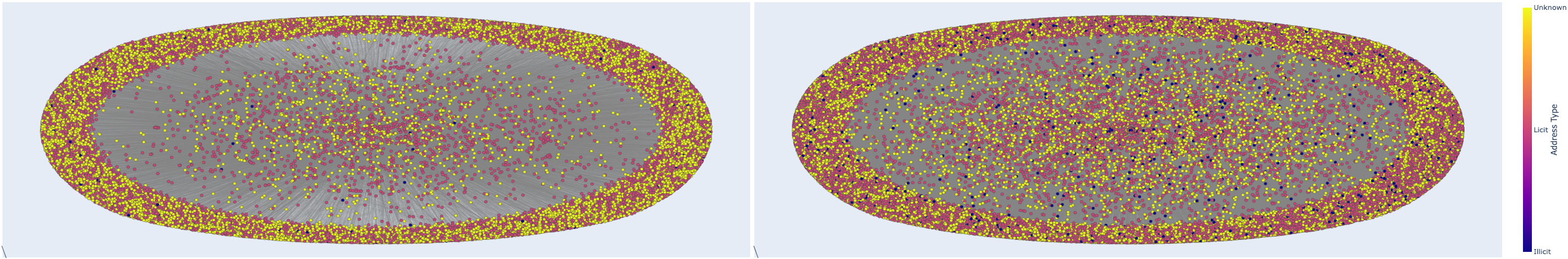}}
\label{fig:walletstimestep1432}
}
\vspace{-0.7cm}
\end{figure*} 

\begin{figure*}
\centering{
\caption{Distribution of transactions and wallet addresses in time steps $14$ (left) and $32$ (right).}\vspace{0.3cm}
\makebox[\linewidth]{\includegraphics[width=7in]{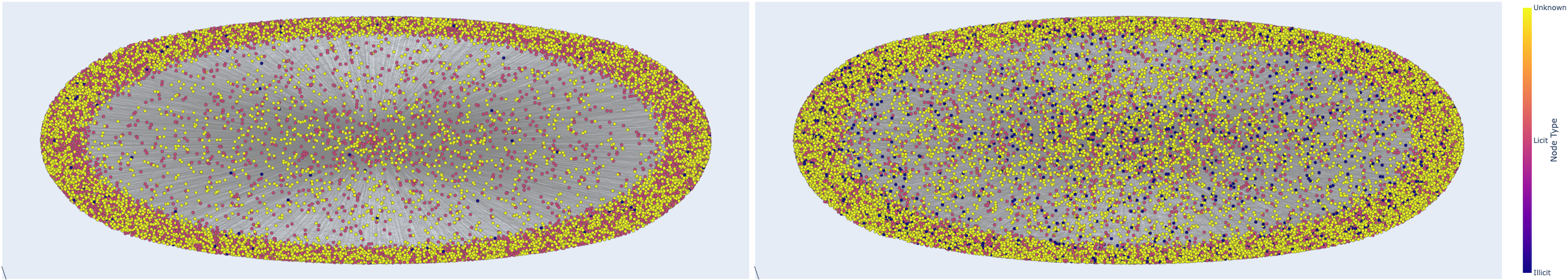}}
\label{fig:txswalletstimestep1432}
}
\vspace{-0.3cm}
\end{figure*} 

\subsection{Fraud Detection Evaluation Metrics} \label{metricsappendix}
The accuracy of the models is verified using five metrics: Precision, Recall, F1 Score, Micro-Avg F1 Score, and Matthews Correlation Coefficient. The explanations have the following components: true positive (TP), false positive (FP), false negative (FN), and true negative (TN).

\begin{enumerate}
    \item \textit{Precision} - the fraction of positive points that were correctly classified. It is calculated by: \\  $$Precision = \frac{TP}{TP + FP}$$\\
    
    \item \textit{Recall} - the fraction of actual positive points that were correctly classified. It is calculated by: \\ $$Recall = \frac{TP}{TP + FN}$$\\
    
    \item \textit{F1 Score} - the harmonic mean of precision and recall. It is calculated by: \\
    $$F1 = 2\cdot \frac{precision\cdot recall}{precision + recall}$$\\

    \item \textit{Micro-Avg F1 Score} - similar to F1 Score, but using the total number of TP, FP, and FN instead of individually for each class. It is calculated by:\\
    $$MicroAvg\;\; F1 = \frac{TP}{TP + \frac{1}{2}\cdot(FP + FN)}$$\\

    \item \textit{Matthews Correlation Coefficient (MCC)} - measures the quality of binary classifications, taking into account the difference between predicted values and actual values. It ranges from $[-1,\;+1]$ with the extreme values $-1$ and $+1$ in cases of perfect misclassification and perfect classification respectively. It is calculated by: \\ \\$$MCC = \frac{(TP \cdot TN)-(FP \cdot FN)}{\sqrt{(TP + FP) \cdot (TP + FN) \cdot (TN + FP) \cdot (TN + FN)}}$$\\
\end{enumerate}


\subsection{Graph Visualization}
Figures~\ref{fig:txstimestep1432},~\ref{fig:walletstimestep1432}, and~\ref{fig:txswalletstimestep1432} show the distributions of transactions and/or wallet addresses for the Money Flow Transaction Graph, the Actor Interaction Graph, and the Address-Transaction Graph respectively. Specifically, they show a comparison between distributions in time step $14$, which produce more sparse graphs, and time step $32$, which produce more dense graphs. The grey area is due to the density of the edges.

\begin{figure*}
\centering{
\caption{Clustering Bitcoin addresses: (a) collect the set of input addresses for each transaction in the address-transaction graph; (b) group transactions with $\geq1$ common element in each set into unique users; (c) highlight transactions within each user in the original graph; and (d) convert highlighted address-transaction graph into a user graph, with edges across users if a path existed between groups of transactions in the original graph.}
\vspace{0.5cm}
\makebox[\linewidth]{\includegraphics[width=7.8in]{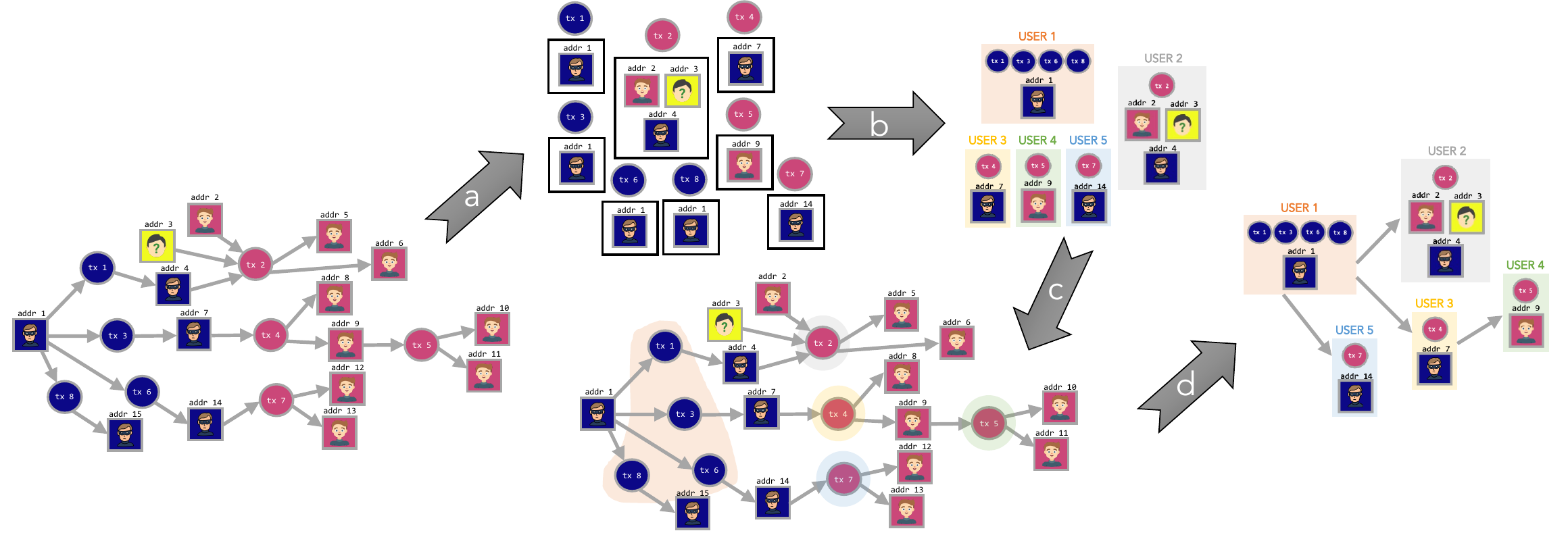}}
\label{fig:fullclusters}
\vspace{0.7cm}
}
\end{figure*}

\subsection{Bitcoin Address Clustering} \label{clusteringappendix}
\begin{figure}[t]
\vspace{-0.4cm}
\centering{
\caption{Bitcoin address clustering based on multiple input addresses heuristic in an address-transaction graph.}
\vspace{0.3cm}\makebox[\linewidth]{\includegraphics[width=3.7in]{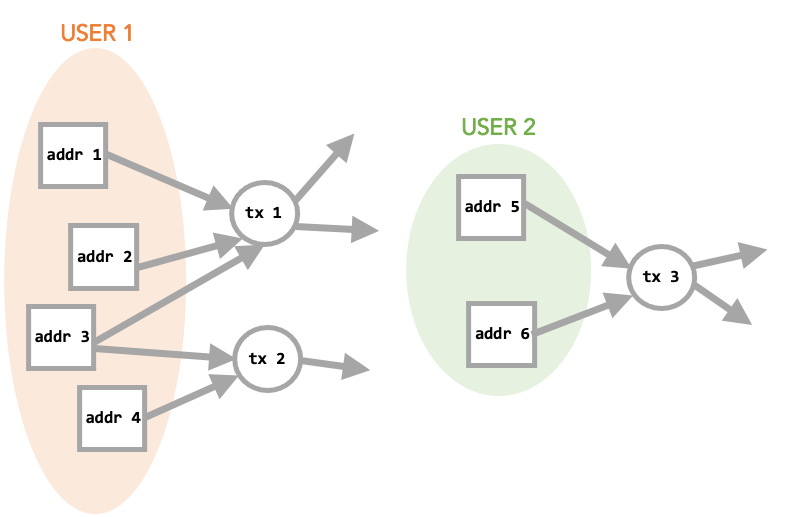}}
\label{fig:clusteringprocess}
}
\end{figure}

The Bitcoin address clustering process discussed in this study is referred to as the \textit{multiple input addresses heuristic}. This heuristic takes advantage of the fact that in order to transact correctly, a user must provide the signatures corresponding to the private keys and the public key wallet addresses for all the input addresses, thus, a user must have access over all private keys. Hence, allowing for the assumption that all input addresses of a transaction can be attributed to the same user. This is demonstrated in Figure~\ref{fig:clusteringprocess}, where the input addresses \texttt{addr1}, \texttt{addr2}, \texttt{addr3}, and \texttt{addr4} are grouped into User $U_1$, while input addresses \texttt{addr5} and \texttt{addr6} are grouped into User $U_2$. Figure~\ref{fig:fullclusters} shows the address clustering process using the four steps discussed in Section~\ref{userentity}, producing a user graph on the right.

\subsection{Statistical Analysis of the Dataset} \label{statsanalysisappendix}

\begin{figure*}[bp!]
\centering{
\caption{Number of transactions for illicit actors appearing in only $1$ time step. Each point is a unique illicit actor. Y-axes are in $\log$ scale.}
\includegraphics[width=7in]{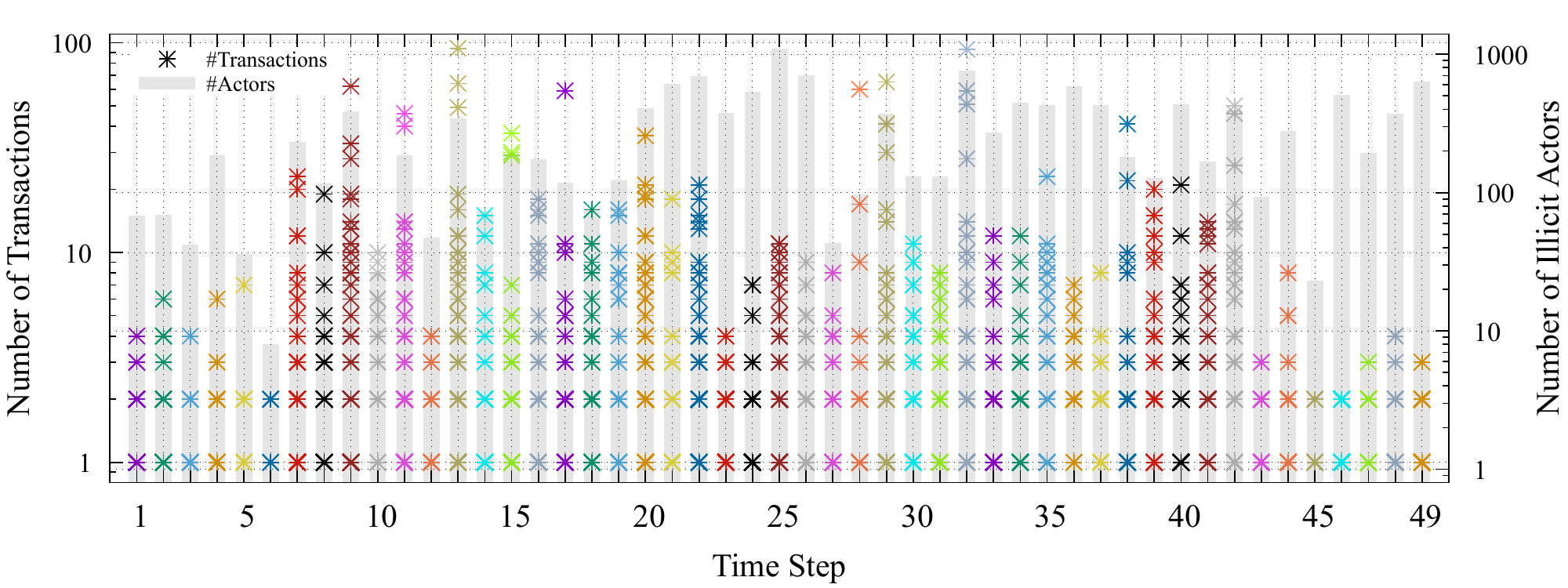}
\label{fig:1000badactor}
}
\end{figure*}

Figure~\ref{fig:1000badactor} shows illicit actors that exist in only $1$ time step ($14,007$ total illicit actors). Aside from providing information about the timeline of the illicit actors, it displays the distribution of the illicit actors among the dataset for these types of actors. For example, in time step 6 there are 8 illicit actors, in time step 14 there are 74 illicit actors, and in time step 32 there are 759 illicit actors. 

\begin{figure*}
\centering{
\caption{Trend comparison for features in the transactions dataset (left) and actors dataset (right). Green highlight shows good correlation (top 2 rows), while red highlight shows unclear correlation (bottom 2 rows).}\label{fig:featurescombinedgraphappendixnew}\vspace{0.3cm}
\makebox[\linewidth]{\includegraphics[width=8in]{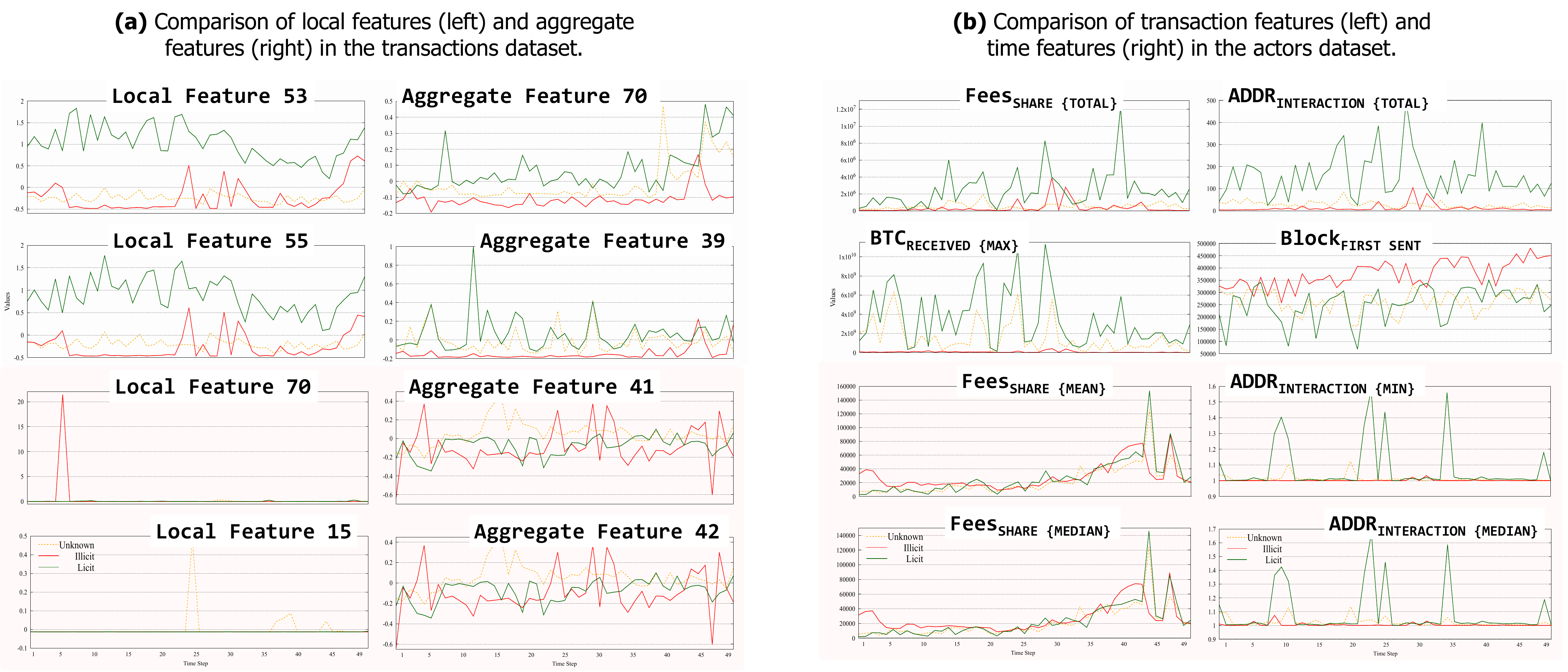}}
}
\end{figure*}

Figure~\ref{fig:featurescombinedgraphappendixnew} is an extension of the dataset feature analysis and feature refinement by displaying the top $2$ and bottom $2$ of each feature class in (a) the transactions dataset (left: local features, right: aggregate features), and (b) the actors dataset (left: transactions features, right: time features). This further solidifies the explainability behind the risk of transactions and actors through the visual (and numerical) distinction among the illicit and licit classes. In all green highlighted figures, one can clearly classify between both classes. Though, this is unlike the red highlighted figures where there is no characteristic trend among both classes.

\subsection{Data Distribution}
Tables~\ref{tab:txspertimesteptable} and~\ref{tab:walletspertimesteptable} show the numerical distribution 
within each class by time step for the transactions and actors datasets respectively.

\clearpage

\begin{table*}[bp!]
\renewcommand{\arraystretch}{1.3}
\centering 
\caption{Numerical distribution of the number of transactions of each class by time step. First two tables are for the training set and the third table is for the testing set.}
\label{tab:txspertimesteptable}
\begin{tabular}{|c||c|c|c|c|c|c|c|c|c|c|c|c|c|c|c|c|c|}
\hline
\textbf{Time step} & \textbf{1} & \textbf{2} & \textbf{3} & \textbf{4} & \textbf{5} & \textbf{6} & \textbf{7} & \textbf{8} & \textbf{9} & \textbf{10} & \textbf{11} & \textbf{12} & \textbf{13} & \textbf{14} & \textbf{15} & \textbf{16} & \textbf{17}\\
\hline
\hline
Unknown & 5733 & 3427 & 5342 & 4253 & 4921 & 3843 & 4845 & 3292 & 4218 & 5755 & 3600 & 1541 & 3719 & 1605 & 3021 & 2445 & 2574\\
\hline
Illicit & 17 & 18 & 11 & 30 & 8 & 5 & 102 & 67 & 248 & 18 & 131 & 16 & 291 & 43 & 147 & 128 & 99\\ \hline
Licit & 2130 & 1099 & 1268 & 1410 & 1874 & 480 & 1101 & 1098 & 530 & 954 & 565 & 490 & 518 & 374 & 471 & 402 & 712\\ \hline
\end{tabular}
\end{table*}
\addtocounter{table}{-1} 
\begin{table*}[bp!]
\vspace{-0.2cm}
\renewcommand{\arraystretch}{1.3}
\centering
\begin{tabular}{|c||c|c|c|c|c|c|c|c|c|c|c|c|c|c|c|c|c|}
\hline
\textbf{Time step} & \textbf{18} & \textbf{19} & \textbf{20} & \textbf{21} & \textbf{22} & \textbf{23} & \textbf{24} & \textbf{25} & \textbf{26} & \textbf{27} & \textbf{28} & \textbf{29} & \textbf{30} & \textbf{31} & \textbf{32} &\textbf{33} & \textbf{34}\\
\hline
\hline
Unknown & 1587 & 2761 & 3391 & 2896 & 4131 & 2978 & 3466 & 1720 & 2006 & 883 & 1369 & 3101 & 1959 & 2106 & 3202 & 2710 & 1971\\
\hline
Illicit & 52 & 80 & 260 & 100 & 158 & 53 & 137 & 118 & 96 & 24 & 85 & 329 & 83 & 106 & 342 & 23 & 37\\ \hline
Licit & 337 & 665 & 640 & 541 & 1605 & 1134 & 989 & 476 & 421 & 182 & 199 & 845 & 441 & 604 & 981 & 418 & 478\\ \hline
\end{tabular}
\end{table*}
\addtocounter{table}{-1} 
\begin{table*}[bp!]
\vspace{-0.2cm}
\renewcommand{\arraystretch}{1.3}
\centering
\begin{tabular}{|c||c|c|c|c|c|c|c|c|c|c|c|c|c|c|c|c|c|}
\hline
\textbf{Time step} & \textbf{35} & \textbf{36} & \textbf{37} & \textbf{38} & \textbf{39} & \textbf{40} & \textbf{41} & \textbf{42} & \textbf{43} & \textbf{44} & \textbf{45} & \textbf{46} & \textbf{47} & \textbf{48} & \textbf{49}\\
\hline
\hline
Unknown  & 4166 & 4685 & 2808 & 2135 & 1577 & 3270 & 4210 & 4986 & 3693 & 3384 & 4377 & 2807 & 4275 & 2483 & 1978\\
\hline
Illicit & 182 & 33 & 40 & 111 & 81 & 112 & 116 & 239 & 24 & 24 & 5 & 2 & 22 & 36 & 56\\ \hline
Licit & 1159 & 1675 & 458 & 645 & 1102 & 1099 & 1016 & 1915 & 1346 & 1567 & 1216 & 710 & 824 & 435 & 420\\ \hline
\end{tabular}
\end{table*}\addtocounter{table}{2}

\begin{table*}[bp!]
\renewcommand{\arraystretch}{1.3}
\vspace{1cm}
\centering 
\caption{Numerical distribution of the number of wallet addresses of each class by time step. First two tables are for the training set and the third table is for the testing set.}
\label{tab:walletspertimesteptable}
\makebox[\linewidth]{
\begin{tabular}{|c||c|c|c|c|c|c|c|c|c|c|c|c|c|c|c|c|c|}
\hline
\textbf{Time step} & \textbf{1} & \textbf{2} & \textbf{3} & \textbf{4} & \textbf{5} & \textbf{6} & \textbf{7} & \textbf{8} & \textbf{9} & \textbf{10} & \textbf{11} & \textbf{12} & \textbf{13} & \textbf{14} & \textbf{15} & \textbf{16} & \textbf{17}\\
\hline
\hline
Unknown & 26465 & 15878 & 26213 & 31035 & 24621 & 17029 & 23357 & 17631 & 23819 & 38271 & 18250 & 10343 & 26813 & 5912 & 11500 & 9064 & 17066\\
\hline
Illicit & 96 & 99 & 58 & 266 & 54 & 17 & 463 & 267 & 960 & 120 & 494 & 77 & 918 & 156 & 447 & 427 & 321\\ \hline
Licit & 24547 & 16292 & 4835 & 8647 & 10465 & 1957 & 7902 & 16229 & 3894 & 8454 & 10576 & 2875 & 3288 & 1941 & 6671 & 2315 & 4039\\ \hline
\end{tabular}
}
\end{table*}
\addtocounter{table}{-1} 
\begin{table*}[bp!]
\renewcommand{\arraystretch}{1.3}
\centering
\makebox[\linewidth]{
\begin{tabular}{|c||c|c|c|c|c|c|c|c|c|c|c|c|c|c|c|c|c|}
\hline
\textbf{Time step} & \textbf{18} & \textbf{19} & \textbf{20} & \textbf{21} & \textbf{22} & \textbf{23} & \textbf{24} & \textbf{25} & \textbf{26} & \textbf{27} & \textbf{28} & \textbf{29} & \textbf{30} & \textbf{31} & \textbf{32} &\textbf{33} & \textbf{34}\\
\hline
\hline
Unknown & 6413 & 11633 & 32483 & 21725 & 28827 & 14101 & 19396 & 12231 & 16434 & 9130 & 6489 & 17180 & 8882 & 13328 & 15679 & 10109 & 9914\\
\hline
Illicit & 189 & 320 & 950 & 1256 & 1491 & 734 & 1083 & 2243 & 1334 & 89 & 219 & 979 & 287 & 312 & 1730 & 561 & 908\\ \hline
Licit & 1643 & 3220 & 6315 & 16576 & 9693 & 7301 & 3518 & 8314 & 2360 & 1000 & 920 & 3811 & 2931 & 5119 & 4873 & 7675 & 5546\\ \hline
\end{tabular}
}
\end{table*}
\addtocounter{table}{-1} 
\begin{table*}[bp!]
\renewcommand{\arraystretch}{1.3}
\centering
\makebox[\linewidth]{
\begin{tabular}{|c||c|c|c|c|c|c|c|c|c|c|c|c|c|c|c|c|c|}
\hline
\textbf{Time step} & \textbf{35} & \textbf{36} & \textbf{37} & \textbf{38} & \textbf{39} & \textbf{40} & \textbf{41} & \textbf{42} & \textbf{43} & \textbf{44} & \textbf{45} & \textbf{46} & \textbf{47} & \textbf{48} & \textbf{49}\\
\hline
\hline
Unknown & 23898 & 34887 & 16578 & 16820 & 9679 & 23564 & 22123 & 31496 & 24953 & 16255 & 22201 & 17719 & 19726 & 13638 & 10030\\
\hline
Illicit & 943 & 1226 & 804 & 400 & 282 & 801 & 454 & 967 & 156 & 497 & 42 & 626 & 256 & 433 & 789\\ \hline
Licit & 6229 & 11130 & 3455 & 4039 & 4881 & 10233 & 7033 & 11142 & 7302 & 10631 & 11057 & 4723 & 8106 & 8783 & 4385\\ \hline
\end{tabular}
}
\end{table*}\addtocounter{table}{2}

\end{document}